\documentclass[preprint,nofootinbib,aps,superscriptaddress,floatfix]{revtex4}
\usepackage{graphicx}
\usepackage{bm}
\usepackage{epsfig}

\newif\ifpdf
\ifx\pdfoutput\undefined
\pdffalse 
\else
\pdfoutput=1 
\pdftrue
\fi

\def\OMIT#1{}

\newcommand{\nn}{\nonumber}

\newcommand{\bea}{\begin{eqnarray}}
\newcommand{\eea}{\end{eqnarray}}

\newcommand{\LQCD}{{\Lambda_{\rm QCD}}}

\newcommand{\Bdecay}{$ B \to X_c \,  \ell \,  \bar{\nu} $ }
\newcommand{\orderalpha}{$ \mathcal{O}\!\left(\alpha_s \right)  $}
\newcommand{\orderalphalam}{$ \mathcal{O}\!\left( \alpha_s \, \LQCD / m_b
\right)  $ }
\newcommand{\orderalphalamsec}{$ \mathcal{O}\!\left( \alpha_s \, \Lambda_{\rm
QCD}^2 / {m_b^2} \right)  $ }
\newcommand{\orderlamsec}{$ \mathcal{O}\! \left( \Lambda_{\rm QCD}^2  /
{m_b^2}  \right) $  }
\newcommand{\orderlamthr}{$ \mathcal{O}\! \left(  \Lambda_{\rm QCD}^3  /
{m_b^3}  \right) $}
\newcommand{\elmin}{E_\ell^{\rm{min}}}

\begin{document}
\ifpdf
\DeclareGraphicsExtensions{.pdf, .jpg}
\else
\DeclareGraphicsExtensions{.eps, .jpg}
\fi

\preprint{  \hbox{hep-ph/0402120}  }
\vspace{0.25cm}
\title{ Improving Extractions of $|V_{cb}|$ and $m_b$ from the
Hadronic Invariant Mass Moments of Semileptonic Inclusive $B$ decay }
\vspace{0.25cm}
\author{Michael Trott}
\affiliation{Department of Physics, University of Toronto, \\ 60 St. George St.
	Toronto, Ontario, Canada, M5S 1A7 \footnote{Electronic address:
mrtrott@physics.utoronto.ca}}
\vspace{0.25cm}
\date{\today\\ \vspace{1cm}}


\begin{abstract}
 We calculate the hadronic tensor for inclusive semileptonic \Bdecay decay to
\orderalpha. This allows  \orderalphalam
 corrections to hadronic invariant mass observables
to be directly evaluated with experimentally required cuts on phase space.
Several moments of phenomenological interest are presented to order
\orderalphalam and
\orderlamthr, allowing a consistent extraction of the HQET parameters up to
\orderlamthr $\,$
and the $b$ quark mass with theoretical error $\sim  50 \, { \rm{{\rm MeV}}} \,
$.
The hadronic invariant mass spectrum is  examined with a general moment to
obtain observables that
test the theoretical error estimate assigned to these parameters;
in particular, fractional moments that directly test the OPE for
inconsistencies in the hadronic invariant mass
spectrum are reported. The $m_b \, \LQCD/m_c^2$ expansion present for 
fractional moments of the hadronic invariant mass spectrum is discussed
and shown to introduce a numerically suppressed uncertainty of $\mathcal{O}\!
\left( m_b ^4 \, \Lambda_{\rm QCD}^4\ / m_c^8 \right)$.

\end{abstract}

\maketitle


\section{Introduction}
Inclusive semileptonic \Bdecay decay offers an opportunity to measure the
Cabbibo-Kobayashi-Maskawa (CKM) parameter $|V_{cb}|$ and the bottom quark mass
\cite{ope, opecorr1,opecorr2,leptonmomentValoshin,
leptonmoment4,leptonmoment2,hadronmoment2,hadronmoment3,hadron3rdorder}.
Measurements of these parameters are crucial to the $B$ factory program of
over-constraining the CKM sector of the
standard model \cite{Falkreview}. Experimental studies of moments of the
differential decay spectrum of \Bdecay combined with a measurement
of the total inclusive decay rate are useful in extracting these parameters, as
these observables can be measured cleanly by experiment, and calculated
from QCD without model dependence using an operator product expansion (OPE).

The OPE is an expansion in powers of the ratio $\LQCD / { m_b } $, where the
terms in this expansion are parameterized
using heavy quark effective theory (HQET).
The OPE demonstrates that in the $m_b \rightarrow \infty$ limit, inclusive $B$
meson decay spectra are equal to $b$ quark decay spectra.
To extract $m_b$ and $|V_{cb}|$ from the inclusive decay spectrum with high
precision, one needs to accurately know the
relevant matrix elements of terms in the OPE that the spectrum depends upon.
Extensive theoretical effort has been devoted to calculating
the decay rates and moments of various spectra; to test HQET by the extractions
 of
these nonperturbative parameters from different spectra, and to obtain
$|V_{cb}|$ and a precise value
for the $b$ quark mass.

Experimental results have been reported by the BABAR, CLEO and DELPHI
collaborations measuring
various $B$ meson decay spectra and moments
\cite{CLEO2001,CLEO20012,CLEO2002,BABAR01,DELPHI02604,DELPHI02605}.
Recent analysis of this data \cite{everyonebutme,russianfit}
finds $|V_{cb}| = \left(40.8 \pm 0.9 \right) \times 10^{-3}$ and $m_{b}^{1S} =
4.74 \pm 0.10 \, \rm{ GeV}$,
where the experimental uncertainties dominate the extraction. As the
experimental errors are
expected to decrease in the near future, it is appropriate to reexamine the
theoretical
error assigned in this extraction.
The largest contributions to the theoretical uncertainties introduced in these
fits come from
the estimated size of the \orderalphalam  corrections with
a lepton energy cut,  the \orderalphalamsec terms,
and the $ \mathcal{O}\! \left(\Lambda_{\rm QCD}^4\ / m_b^2 m_c^2 \right)$ terms
introduced due to the HQET
expansion employed  for $m_c$ \cite{uraltsev}.

In past calculations, the lepton energy cut dependence of the \orderalphalam
terms was not calculated,
and these terms were treated as a source of error in the determination of $m_b$
and $|V_{cb}|$.
In this paper, we improve upon past results by calculating the lepton energy
cut
dependence of the $\mathcal{O}\! \left(\alpha_s \, \Lambda_{\rm QCD}  /  m_b
\right)$ terms.
With the calculation of these terms and the moments presented in this paper,
global fits will allow precise
determinations of $|V_{cb}|$ and $m_b$ to occur from the inclusive decay
spectrum.

As the precision of determinations of $m_b$ and $|V_{cb}|$ improves,
it becomes important to test the consistency of the OPE more precisely.
Observables that do not depend strongly on the nonperturbative parameters that
introduce the
dominant uncertainty in extractions of $m_b$ and $|V_{cb}|$ allow one to test
if the uncertainty
assigned for all higher order terms in the OPE is sufficiently large.
By examining a general moment of the lepton spectrum \cite{christianmike},
observables of this type, called OPE testing moments, have been found.
In this paper, we apply this technique to hadronic invariant mass moments.
By testing the error assigned to higher order effects experimentally
we improve the confidence in the theoretical error assigned due to these
effects in determinations of $m_b$ and $|V_{cb}|$ from moments of semileptonic 
inclusive b decay.
This allows extractions of $m_b$ from these decay to occur in
a relatively theoretically clean and
unambiguous fashion \cite{lukebmass}. These results can be combined with the
results for the lepton energy spectrum for
cross checks and fits to determine the HQET parameters.

The structure of this paper is as follows. In Section \ref{tensor} the
$\mathcal{O}\! \left(\alpha_s \right)$
contribution to the hadronic tensor is presented.
Section \ref{moments} reports on moments in the 1S  mass scheme
\cite{renorupsilon,upmassmanohar,renormalon,luke_renormalon},
and discusses the $m_b  \, \LQCD / m_c $ expansion present in fractional 
hadronic invariant mass moments. The
decay width to \orderalphalam $\,$ and $\mathcal{O}\! \left( \Lambda_{\rm
QCD}^3  /  {m_b^3} \right) $ and the error
that should be assigned in the fit of the moments presented in this paper is
discussed.
The dominant parameters affecting the extraction of $|V_{cb}|$
inclusively,  $m_b^{\rm 1S}$ and $ \lambda_1$, are extracted from known
moments.
Observables appropriate to precisely test the consistency of the OPE are
reported and
moments that allow a measurement of the $b$ quark mass with minimal 
theoretical error due to unknown matrix elements are presented.

\section{\orderalpha $\,$  Contribution to Decay Spectrum}\label{tensor}
\subsection{ Hadron Tensor Decomposition }
The \orderalpha $\,$ corrections to semileptonic \Bdecay  decay have been known
for particular spectra and moments for some time
\cite{jezabekandkhun,corbospec,hadronmoment1}.
The decomposition of the triple differential decay spectrum in terms of
structure functions has not appeared in the literature
to date, although the limit of this spectrum appropriate for a massless final
state is known \cite{Neubertbupaper}.
The triple differential decay spectrum must be known to allow for the
experimentally required cuts on the
kinematic variables to be imposed in calculating the \orderalphalam terms
and to perform a general moment analysis, and so we present it here.

We decompose the triple differential decay spectrum in terms of
the invariant mass of the $W$ boson $\hat{y} = q^2 / {m_b}^2 $ where 
$q^{\mu}$ is the momentum of the lepton pair, the $c$ quark
jet invariant mass $\hat{z} = {\left( m_b \, v - q \right)}^2 / {m_b}^2$,
and the charged lepton energy  $\hat{E}_{\ell} = E_{\ell}/m_b $.
This spectrum is written in terms of a lepton tensor $L_{\mu \nu}$ and the
hadron tensor $W^{\mu \nu}$,
\begin{eqnarray}
\frac{1}{\Gamma_0} \frac{d \Gamma}{d \hat{y} \, d \hat{z}  \,d \hat{E}_{\ell}}
&=&  W^{\mu \nu}(\hat{y} , \hat{z}) L_{\mu \nu}(\hat{y} ,\hat{z},
\hat{E}_{\ell}),
\end{eqnarray}
where
\begin{eqnarray}
\Gamma_0 = \frac{G_F^2 {|V_{cb}|}^2 \left(m_b^{{\rm{pole}}}\right)^5}{192
\pi^3}.
\end{eqnarray}

Integrating over the charged lepton energy the differential decay spectrum
becomes
\vspace{.2cm}
\begin{eqnarray}\label{contraction}
 \frac{1}{\Gamma_0}\frac{d \Gamma}{d \hat{y} \,  d \hat{z} }
 & = & 12\,  E_0 \, t\,  W_{\mu \nu} \left( \hat{y}, \hat{z} \right) L^{\mu
\nu}\left(\hat{y} ,\hat{z} \right) \, ,
\end{eqnarray}

\noindent{where $E_0 = 1/2 \left( 1 + \hat{z} - \hat{y} \right) $ is the
leading order energy of the $c$ quark jet, $\rho = m_c^2/m_b^2 $
and  $t = \sqrt{1 - \hat{z} /{E_0}^2}$} is the rapidity of the $c$ quark.

The hadron tensor can be decomposed in terms of the
initial $B$ meson four momentum $Q^{\mu} $ and the hadronic decay products four
momentum $P^{\mu} = Q^{\mu} - q^{\mu}$,
where $q^{\mu}$ is the momentum of the lepton pair. This tensor can be
calculated from the discontinuity of the time ordered
product of the current $J_{\mu} = \bar{c} \gamma_{\mu}( 1 - \gamma_{5}) b$,
\begin{eqnarray}
W^{\mu \nu} =  \frac{1}{\pi \, \Gamma_0 } \mbox{Im}\left[i\int d^4 x e^{i(P -
m_b v)\cdot x} \langle\bar{B}|T[J^{\dag \mu}(x), J^{\nu}(0)]|\bar{B}\rangle
\right].
\end{eqnarray}

This tensor is calculated by considering the quark-gluon level processes
involved in this decay.
The spectra obtained from the parton level discontinuity are expected to
accurately describe physical $B$ meson decay spectra
so long as observables are sufficiently inclusive.

The tensor decomposition in terms of the four vectors $Q^{\mu}$ and $P^{\mu}$
yields five non trivial structure functions ${W_i}$,
\begin{eqnarray}\label{structurefunction}
\! W^{\mu \nu}(\hat{y},\hat{z}) &=&W_1(\hat{y},\hat{z}) \left(P^{\mu}Q^{\nu}+
P^{\nu} Q^{\mu} - P \cdot Q g^{\mu \nu} +
i \epsilon^{\mu \nu \alpha \beta} Q_{\alpha} P_{\beta} \right) \\
&-&  W_2 (\hat{y} , \hat{z}) g^{\mu \nu} + W_3 (\hat{y} ,\hat{z}) Q^{\mu}
Q^{\nu}
+ W_4 (\hat{y} , \hat{z}) ( P^{\mu} Q^{\nu} + P^{\nu} Q^{\mu}) +W_5 (\hat{y} ,
\hat{z})  P^{\mu} P^{\nu}. \nonumber
\end{eqnarray}

The operator product expansion of the structure functions is known to
\orderlamthr \cite{ope, opecorr1,opecorr2,hadron3rdorder}.
There are two nonperturbative parameters at \orderlamsec labelled
$\lambda_{1,2}$ and six parameters at order
\orderlamthr, labelled $\tau_{1,2,3,4}$ and $\rho_{1,2}$ the definitions of
which can be found in \cite{hadron3rdorder}.

\subsection{ \orderalpha $\,$ Contributions to Hadron Tensor Structure
Functions} \label{orderalpha}

The hadronic structure functions  ${W_i}$ in Eq.\ \,(\ref{structurefunction})
have the perturbative expansion:

\begin{eqnarray}
W_i \left( \hat{y}, \hat{z} \right) =
{W_i^{0}} \left( \hat{y}, \hat{z} \right) + \frac{C_f \, \alpha_s}{4 \, \pi}
{W_i^1} \left( \hat{y}, \hat{z} \right) + \mathcal{O}\! \left( {\alpha_s^2} \right).
\end{eqnarray}

\begin{figure}
\includegraphics[width=9.5cm]{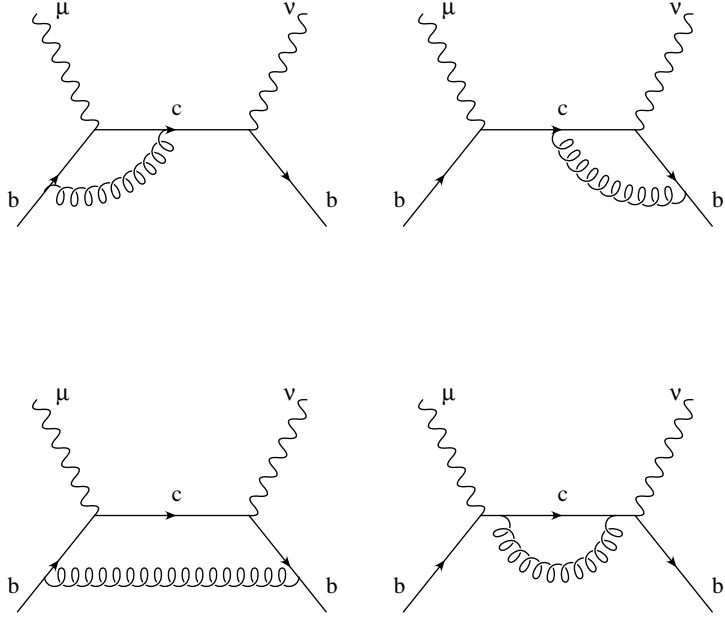}
\caption{The one loop forward scattering diagrams. The hadron tensor is derived
by calculating the imaginary part of the diagrams.\label{fig1}}
\end{figure}

The $\mathcal{O}\! \left(\alpha_s \right)$ contributions to the structure functions 
are calculated by taking all cuts across all intermediate state
contributions to the diagrams shown in Fig.~\ref{fig1}. Combined with the
external quark wave-function
renormalization this gives the  $W_i^i$. At tree level, in the massless final
state limit only $W_1^0$ is nonzero.
$W_1^1$ can be expressed as
\begin{eqnarray}
 W_1^1(\hat{y} , \hat{z})  &=&    W_{1a}^1 (E_0,t, \hat{z},\hat{y}) \, \delta
(\hat{z} - \rho )
  + \, W_{1b}^1 (  E_0,t,\hat{z})\,  \theta (\hat{z} - \rho ) +
W_{1c}^1 ( E_0,t,\hat{z}, {\lambda_G}^2),
\end{eqnarray}
where
\begin{eqnarray}
W_{1a}^1 (E_0,t, \hat{z},\hat{y}) &=& - 16
- \frac{2}{t} \log^2 \left( \frac{1 + t}{1 - t}\right)  - \frac{4}{t} \log
\left( \frac{1 + t}{1 - t}\right) \log \left( \rho \right) \nonumber \\
&+& \left( \frac{4 \, \left( 1 + \rho \right)}
{\hat{y} \, t} - \frac{12 \, E_0}{\hat{y}\, t} +  \frac{4 \, \rho}
{\hat{y}\, E_0 \, t}  \right) \log \left( \frac{1 + t}{1 - t}\right)
+\frac{ \left(2 - 8 E_0 + 6 \, \rho \right)}
{\hat{y}} \log \left( \rho \right)
\nonumber  \\
&+& \frac{8}{t} \,  {\rm Li}_2 \left(\frac{2\, E_0  \, t}
{ E_0 \left( t -1 \right) + 1} \right) 
- \frac{8}{t} \, {\rm Li}_2 \left(\frac{2\, E_0  \, t}{ E_0 \left(1+ t
\right) - \rho}\right)  \nonumber \\
&-& \frac{8}{t}\left( \log
\left( \frac{1 + t}{1 - t}\right) + \log \left( \rho \right) \right) 
\log\left( \frac{1 + \left(t - 1 \right) {E_0}}
{\sqrt{\hat{y}}} \right) ,\, \nonumber \\
\nonumber \\
W_{1b}^1 (E_0,t, \hat{z}) &=& \frac{-2\,\left( 1 + \hat{z} \right) \,\left(
\hat{z} - \rho  \right) }{{\hat{z} \,{E_0}^2}\,{t^2}} +
 {\frac{2\,\left( 5\,\hat{z} - \rho  \right) }{\hat{z} \, {E_0}\,{t^2}}}
- {\frac{2\,\left( 3\,\hat{z} + \rho  \right) }{{\hat{z}^2} \,{t^2}}}
 \, \nonumber \\
&\,& - \left( \frac{4}{ E_0 \,  t^2 } - \frac{ \left(\hat{z} - \rho + 4
\right)}{{E_0}^2 \, t^2} + \frac{ \left(\hat{z} - \rho \right)}{{E_0}^3\, t^2}
\right)
\frac{1}{t} \log \left( \frac{1 + t}{1 - t}\right)
, \, \nonumber \\
\nonumber \\
W_{1c}^1 ( E_0,t,\hat{z}, {\lambda_G}^2) &=&  - \left( 16 - \frac{8}{t} \log
(\frac{1 + t}{1 - t})  \right) \log (\lambda_G ) \delta (\hat{z} - \rho )
\nonumber \\
&-& \theta{\left( \hat{z} - \left( \sqrt{\rho} + \lambda_{G}\right)^2 \right)}
\frac{8\,\rho \, f_1}{\hat{z}\, {\left( \hat{z}  - \rho + {\lambda_G}^2
\right)}^2}
- \theta{\left( \hat{z} - \left( \sqrt{\rho} + \lambda_{G}\right)^2 \right)}
\frac{8\, f_1}{f_2} \nonumber \\
&+& \theta{\left( \hat{z} - \left( \sqrt{\rho} + \lambda_{G}\right)^2 \right)}
\frac{8\,\log \left(\frac{ \hat{z} + {\lambda_G}^2 - \rho +
f_1 \,t  }
{ \hat{z} + {\lambda_G}^2 - \rho -  f_1 \,t }  \right)}
{t\,\left( \hat{z}  - \rho + {\lambda_G}^2 \right)}, \nonumber
\end{eqnarray}
and
\begin{eqnarray}
  f_1 &=&  \sqrt{{\hat{z}^2} + {\left( {\lambda_G}^2 - \rho  \right) }^2 -
2\,\hat{z}\,\left( {\lambda_G}^2 + \rho  \right) }, \nonumber \\
  f_2 &=& {\hat{z}^2} + 2\,{E_0}^2 \,\left( 1 + {t^2} \right) \,{\lambda_G}^2 +
 {\left( {\lambda_G}^2 - \rho  \right) }^2 - 2\,\hat{z}\,\rho. 
\end{eqnarray}

\vspace{0.1cm}
\noindent{The IR singularities present in the unintegrated spectra are
regulated by a
gluon mass $\lambda_G$ in this calculation and the divergence cancels between
the virtual
and bremsstrahlung graphs once one integrates over phase space.
The divergence directly cancels in integrations of these structure functions as
the first two bremsstrahlung terms in $W_{1c}^1 $ each contribute
a factor of $ \rm{8\, log \left(\lambda_G \right)}$, while the final term
contributes a factor of
$\rm{8/t \, log \left(\frac{1+t}{1-t} \right) \, log \left(\lambda_G \right)}$.
For the purposes of this paper it is sufficient to numerically integrate the
\orderalpha $\,$ spectrum
with the regulator assigned a small numerical value $\lambda_G  \sim 10^{-6}$.
The $ {\rm Li}_2\left(\hat{z}\right)$ functions in $ W_1^1 $ are the
Dilogarithm functions, defined as
${\rm Li}_2\left(\hat{z}\right) \equiv \sum_{k = 1}^{\infty} \frac{z^k}{k^2}
\,$ or equivalently, as
${\rm Li}_2\left(\hat{z}\right) \equiv \int_z^0 \frac{\log \left(1- t
\right)}{t} dt$.}
The other structure functions vanish at tree level in the limit $m_c
\rightarrow 0$
and are IR safe at \orderalpha. For these structure functions we find,
\newpage
\begin{eqnarray}
 W_2^1(\hat{y} , \hat{z})  &=&     W_{2a}^1 (E_0,t, \hat{z}) \,  \delta
(\hat{z} - \rho )
  +   W_{2b}^1 ( E_0,t,\hat{z})\,  \theta(z - \rho ),  \nonumber \\
 W_{2a}^1 (E_0,t, \hat{z}) &=& - \frac{2 \,\left(\hat{z} + \rho \right)}{E_0 \,
t} \log \left( \frac{1 + t}{1 - t}\right),
\,  \\
W_{2b}^1 ( E_0,t,\hat{z}) &=&  \frac{\left( \hat{z}^2 + 8 \hat{z} - \rho^2
\right)}{\hat{z} \,E_0 \, t^4} - \frac{ 4 \left(\hat{z} + \rho \right)}{{E_0}^2
\, t^4}
- \frac{\left( \hat{z}^2 + 8 \hat{z} - \rho^2 \right)}{{E_0}^3 \, t^4} + \frac{
4 \, \hat{z}\, \left(\hat{z} + \rho \right)}{{E_0}^4 \, t^4}  \nonumber \\
&+& \left( -\frac{4}{E_0} +   \frac{2 \left( \hat{z} + \rho \right)}{{E_0}^2}
- \frac{ \left( \hat{z}^2 - 8 \, \hat{z} - \rho^2 \right)}{2 {E_0}^3}
  - \frac{2 \,\hat{z} \, \left( \hat{z} + \rho \right)}{{E_0}^4 }+ \frac{
\hat{z} \, \left(\hat{z}^2 - \rho^2 \right)}{2 \, {E_0}^5}\right)
  \frac{1}{t^5} \log ({\frac{1 + t}{1 - t}}) \nonumber \\
\, \nonumber \\
 W_3^1(\hat{y} , \hat{z})  &=&  W_{3a}^1 (E_0,t, \hat{y}) \,  \delta (\hat{z} -
\rho ) +   W_{3b}^1 ( E_0,t,\hat{z})\,  \theta(z - \rho )  \nonumber \\
 W_{3a}^1 (E_0,t, \hat{y}) &=&  \frac{\hat{z}}{\hat{y}} \left( \frac{- 2 \,
\left(1 + \hat{y} - \rho \right)}{{E_0}\, t} \, \log \left( \frac{1 + t}{1 - t}
\right)
- 4 \, \log \left( \rho \right) \right)\\
 W_{3b}^1 (E_0,t, \hat{z}) &=&  \frac{16}{{E_0} \, t^4 } -  \frac{ 14 \,\left(
\hat{z} - \rho \right)}{{E_0}^2 \, t^4}
+ \frac{3 {\left( \hat{z} - \rho \right)}^2 - 16 \hat{z}}{{E_0}^3 \, t^4}
- 4 \, \hat{z} \, \frac{\left(\hat{z} - \rho \right)}{{E_0}^4 \, t^4}
\nonumber \\
&+& \left( -\frac{8}{E_0 }  + \frac{4 \left( \hat{z}- \rho \right)}{{E_0}^2} -
\frac{{\left( \hat{z}- \rho \right)}^2 - 8\, \hat{z}}{{E_0}^3}
+ \frac{5 \, \hat{z} \, \left(\hat{z} - \rho \right)}{{E_0}^4} - \frac{\hat{z}
{\left( \hat{z} - \rho\right)}^2}{2 \, {E_0}^5}\right)\frac{1}{t^5}
 \log \left( \frac{1 + t}{1 - t} \right) \nonumber \\
\, \nonumber \\
 W_4^1(\hat{y} , \hat{z})  &=&  W_{4b}^1 ( E_0,t,\hat{z})\, \theta(\hat{z} -
\rho ) \nonumber \\
W_{4b}^1 (E_0,t, \hat{z}) &=&  \frac{2\, \left( \hat{z} - \rho
\right)}{\hat{z} \, {E_0} \, t^4 } -  \frac{{\left(\hat{z} - \rho
\right)}^2}{\hat{z} \, {E_0}^2 \, t^4 }
+ \frac{16\, \left( \hat{z} - \rho  \right)}{ {E_0}^3 \, t^4 }  - \frac{2\,
{\left(\hat{z} - \rho \right)}^2 }{{E_0}^4 \, t^4} \nonumber \\
&+& \left(  \frac{- 7 \left( \hat{z} - \rho \right) }{{E_0}^3}  + \frac{3 \,
{\left( \hat{z} - \rho \right)}^2 }{2\, {E_0}^4  } \,
- \frac{2 \,\hat{z} \, \left(\hat{z} - \rho \right)}{{E_0}^5} \right)
\frac{1}{t^5} \log \left( \frac{1 + t}{1 - t} \right) \\
\, \nonumber \\
 W_5^1(\hat{y} , \hat{z})  &=&  W_{5a}^1 (E_0,t, \hat{y}) \, \delta (\hat{z} -
\rho )
 +   W_{5b}^1 ( E_0,t,\hat{z})  \,  \theta(\hat{z} - \rho )  \nonumber \\
W_{5a}^1 (E_0,t, \hat{y}) &=&  \frac{1}{\hat{y}} \left( \frac{ 2 \, \left(1 -
\hat{y} - \rho \right)}{{E_0}\, t} \, \log \left( \frac{1 + t}{1 - t} \right)
+ 4 \, \log \left( \rho \right) \right)  \nonumber \\
W_{5b}^1 (E_0,t, \hat{z}) &=& \frac{2 \,\left( {\hat{z}^2} - {{\rho }^2}
\right) }{  E_0 \,{\hat{z}^2} \, t^4}
- \frac{ 2 \, \left( 11 \hat{z} - 3 \rho \right)}{\hat{z} \, {E_0}^2 \, t^4}
+ \frac{{\left( \hat{z} - \rho \right)}^2 - 4\, \rho \, \left( \hat{z} - \rho
\right)}{\hat{z}\, {E_0}^3 \, t^4 }
+ \frac{4 \, \left(\hat{z} + 3 \rho \right)}{{E_0}^4 \, t^4} \nonumber \\
&+& \left( \frac{8}{{E_0}^2} - \frac{2\, \left( \hat{z} - \rho
\right)}{{E_0}^3} + \frac{\hat{z} -  9 \, \rho}{{E_0}^4}
+ \frac{ \hat{z}^2 + 2\, \hat{z}\, \rho - 3\, \rho^2}{2\, {E_0}^5}\right)
\frac{1}{t^5} \log \left({\frac{1 + t}{1 - t}} \right).
\end{eqnarray}

We have checked the hadron tensor at \orderalpha $\,$ by integrating our
results
to compare against known \orderalpha $\,$ spectra and agree with
\cite{jezabekandkhun} and the historical
\cite{AliPietarinen}, but disagree, as do these other authors, with
\cite{corbospec}.
We also agree with the total \orderalpha $\,$ contribution to the decay rate in
\cite{Nir}.
The  massless limit of the \orderalpha $\,$ hadron structure functions
has been taken for all regular terms and we find we agree with the regular
terms for a massless final state \cite{Neubertbupaper}.

\newcounter{roman}
\setcounter{roman}{1}
\subsection{Lepton Tensor and Phase Space}
To find moments of the triple differential spectrum one must integrate over
phase space
while imposing the experimentally required cut on the lepton energy.
With no lepton energy cut the phase space is given by the following region
\cite{jezabekandkhun},
referred to as region $R_{\Roman{roman}}$, \addtocounter{roman}{1}
\begin{eqnarray}\label{region1}
\frac{1}{2}( b -  \sqrt{{b^2 - \hat{y}}}) \, \,  \le \,  &\hat{E}_{\ell}& \,
\le  \, \, \frac{1}{2}(b+ \sqrt{b^2- \hat{y}})\, \nonumber \\
\left(\sqrt{\rho}+\lambda_G\right)^{2}  \, \,  \le \,  &\hat{z}& \, \,  \le \,
\left(1-\sqrt{\hat{y}}\right)^{2}  \\
 0 \, \,  \le \,  &\hat{y}& \, \le  \, \left(1-\sqrt{\rho} - \lambda_G
\right)^{2} \, \nonumber
\end{eqnarray}

\noindent{where $b \equiv 1/2 \left(1 - \hat{z} + \hat{y} \right) $ and $\rho =
{m_c^2}/{m_b^2}$ .
Without a lepton energy cut in the phase space, the lepton tensor integrated
over the lepton energy $\hat{E}_{\ell}$ is}
\vspace{.2cm}
\begin{eqnarray}\label{simplelepton}
 L_{\mu \nu}(\hat{y} ,\hat{z}) = \hat{y} / {3} \left( -g_{\mu \nu} +
{\hat{q}_{\mu} \hat{q}_{\nu}} / {\hat{y}} \right).
\end{eqnarray}

When a minimum cut $\elmin \ge   x \,   m_b$ is introduced, the phase space and 
the lepton tensor are modified. We do not repeat the derivation of the lepton 
tensor with a cut here, see \cite{hadronmoment4},
but note that the phase space splits into three regions when a cut is imposed.
These three regions correspond to the partitioning of phase space that occurs
when the electron energy lies below or within the phase space integration range
as shown in Fig. 2.
We only consider cuts below the upper limit of $E_{\ell}$ as given in Eq.\
(\ref{region1}), this corresponds to
only considering cuts where $\elmin \le m_b (1 - \sqrt{\rho})/2 $.

\begin{figure}
\includegraphics[width=16cm]{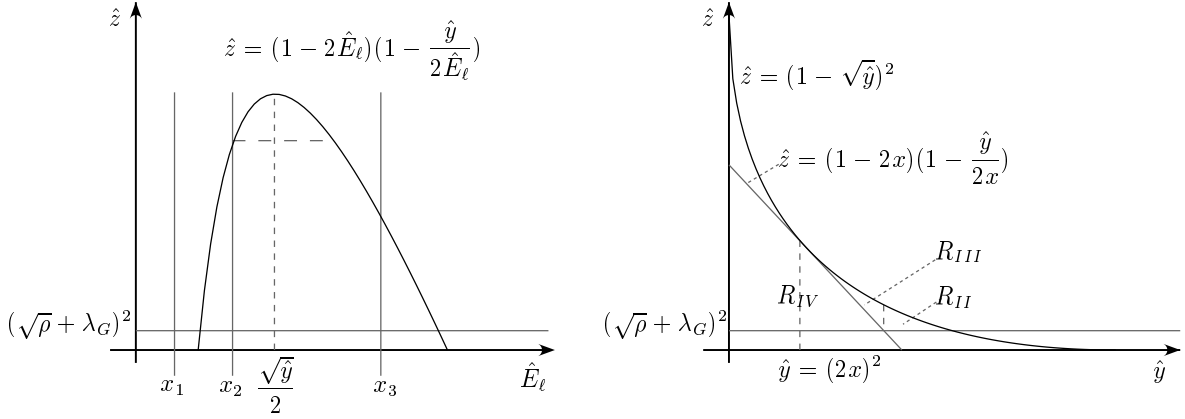}
\caption{Phase space diagrams with a lepton energy cut.\label{figt}}
\end{figure}

For $ x \le 1/2 \left(a - \sqrt{ a^2 -  \hat{y} }\right)$, where $a \equiv 1/2
\left(1 - \left(\sqrt{\rho} + \lambda_G \right)^2 + \hat{y} \right) $
the cut, labelled as $x_1$ in Fig. 2,  is below the lower limit of the 
lepton energy in Eq.\ (\ref{region1}) and as the 
integration over the lepton energy is unaffected, the lepton tensor with this 
cut reduces to the simple expression above with the
electron energy integrated over
\begin{eqnarray}\label{basicEE}
\frac{1}{2}(b- \sqrt{ b^2-\hat{y}}) \le \hat{E}_{\ell} \le \frac{1}{2}(b+
\sqrt{b^2-\hat{y} }).
\end{eqnarray}
\vspace{.2cm}
However, this lepton energy cut still affects the subsequent integrations 
of $\hat{z}$ and $\hat{y}$
by imposing the constraint on the range of $ \hat{y}$ ,
\begin{eqnarray}
( 1 - \left(\sqrt{\rho} + \lambda_G \right)^2 - 2 x) \frac{2 x}{1 - 2x} \, \,
\le \,  \hat{y},
\end{eqnarray}
\noindent{ so that the remaining kinematic variables are integrated over
the phase space region  $R_{\Roman{roman}}$, \addtocounter{roman}{1}}
\begin{eqnarray}
{\left(\sqrt{\rho} + \lambda_G \right)}^2 \, \,  \le \,  &\hat{z}& \, \,  \le
\,  \left(1-\sqrt{\hat{y}}\right)^{2} \,  \\
( 1 - \left(\sqrt{\rho} + \lambda_G \right)^2 - 2 x) \frac{2 x}{1 - 2x} \, \,
\le \,  &\hat{y}& \, \,  \le \,  (1 - \sqrt{\rho}- \lambda_G )^2. \nonumber
\end{eqnarray}
For $ x  \ge 1/2 \left(a - \sqrt{ a^2 -  \hat{y} }\right)$ 
while $x \le  \sqrt{\hat{y}} /2 $,
labelled in the diagram as the cut $x_2$, the phase space splits into two regions.
The first region $R_{\Roman{roman}}$ has the hadron
tensor contracted with the lepton tensor of
Eq.\ \,(\ref{simplelepton}), and the range of $\hat{E}_{\ell}$ as given by Eq.\
\,(\ref{basicEE}).
The remaining phase space variables are then integrated over the range,
\begin{eqnarray}
( 1 - 2 x) ( 1 - \frac{\hat{y}}{2 x}) \, \,  \le \,  &\hat{z}& \, \,  \le \,
\left(1-\sqrt{\hat{y}}\right)^{2} \,  \\
(2 x)^2 \, \,  \le \,  &\hat{y}& \, \,  \le \, ( 1 - \left(\sqrt{\rho} +
\lambda_G \right)^2 - 2 x) \frac{2 x}{1 - 2x}. \, \nonumber
\end{eqnarray}

The second region of type $R_{\Roman{roman}}$\addtocounter{roman}{1} 
combines with the region of  phase space where 
$ x \ge 1/2 \left(a - \sqrt{ a^2 -  \hat{y} }\right)$ 
while $x \ge  \sqrt{\hat{y}} /2 $, labelled as $x_3$ on Fig. 2.
The resulting combined phase space has the lepton tensor incorporating
the cut within the phase space range and the subsequent integration  
is given by the region $R_{\Roman{roman}}$, \addtocounter{roman}{1}
\begin{eqnarray}
x  \, \,  \le \,  &\hat{E}_{\ell}& \, \,  \le \, \frac{1}{2}(b+ \sqrt{ b^2 -
\hat{y}}), \, \nonumber \\
{\left(\sqrt{\rho} + \lambda_G \right)}^2  \, \,  \le \,  &\hat{z}& \, \,  \le
\, ( 1 - 2x)( 1 - \frac{\hat{y}}{2x}) ,\,  \\
0  \, \,  \le \,  &\hat{y}& \, \,  \le \,( 1 - \left(\sqrt{\rho} + \lambda_G
\right)^2 - 2 x) \frac{2 x}{1 - 2x} . \nonumber
\end{eqnarray}
\setcounter{roman}{1}
\section{Hadronic Mass Moments}\label{moments}
\subsection{1S Mass Scheme} \label{upsilon}

In calculating moments of the hadronic invariant mass spectrum, we use the 1S
mass and the upsilon expansion \cite{renorupsilon,upmassmanohar}.
It is well known the pole mass renormalon \cite{renormalon,luke_renormalon}
ambiguity leads to unnecessarily badly behaved perturbation series
for moments when a poor mass scheme is chosen. In the 1S scheme the renormalon
ambiguity is of $\mathcal{O}\!\left(\Lambda^4_{\rm QCD}/m_b^4 \right)$
and so we expect the perturbation series of moments of the spectra to be better
behaved.
We express moments in terms of the 1S mass, which is related to the
$b$ quark pole mass through the relation
\cite{renorupsilon,upmassmanohar}
\begin{eqnarray}\label{masstrans}
\frac{m_b^{\rm 1S}}{ m_b^{\rm pole}} = 1 - \frac{(\alpha_s C_F)^2}{8}
\Big\{1\epsilon
+\frac{\alpha_s}{\pi} \left[ \left(\ell+\frac{11}{6} \right) \beta_0 -4 \right]
\epsilon^2 + \mathcal{O}\! \left( \epsilon^3 \right) \Big\},\end{eqnarray}
where $ \ell = \log[ \mu / (m_{b} \alpha_{s} C_{F} ) ] $ and $C_F = 4/3$.
The parameter $\epsilon = 1$ determines the order in the modified perturbative
expansion.
Using the upsilon expansion necessitates introducing a modified perturbative
expansion in order to ensure the cancellation of renormalon
ambiguities \cite{renorupsilon}. When calculating in the 1S mass scheme the
$\mathcal{O}\!\left({\alpha_{s}}^n \right)$ perturbative
corrections coming from the mass transformation Eq.\ (\ref{masstrans}) are
counted using the parameter $\epsilon^{n-1}$, while
$\mathcal{O}\!\left({\alpha_{s}}^n \right)$
corrections in the decay rate are counted as $\epsilon^{n}$.

The dependence on the pole mass of the charm quark in our results is eliminated
through,
\begin{eqnarray}\label{charmrelation}
m_b^{\rm pole} - m_c^{\rm pole} &=& \bar{m}_B - \bar{m}_D + \lambda_1
\left(\frac{2 \,  \bar m_D -  m_{\Upsilon}}{2 {m}_{\Upsilon} \, \bar m_D}
\right)
+ \lambda_1 \frac{ {m}_{\Upsilon} - 2 \, \bar{m}_B }{4 \bar m_D^2}  + \lambda_1
\Lambda_{1S} \frac{4 \bar m_D^2 -  m_{\Upsilon}^2}{2 \bar{m}_D^2 \,
m_{\Upsilon}^2}
\nonumber \\ \nonumber \\
&-& \left(\tau_1 + \tau_3 - \rho_1 \right)\frac{ m_{\Upsilon}^2 - 4 \bar
m_D^2}{4\bar{m}_D^2 \, m_{\Upsilon}^2}\,
+ \mathcal{O}\! \left(\Lambda_{\rm QCD}^4\ / m_b m_c^2 \right).
\end{eqnarray}
The meson masses $\bar{m}_D$ and $\bar{m}_B$
are the spin averaged meson masses $\bar{m}_X = \left( m_X + 3 m_{X^\ast
}\right)/4$.
In this relation  we use the fact that
$ \frac{m_{\Upsilon}}{2}-m_b^{\rm 1S} \sim \Lambda_{\rm QCD}$ and expand in the
parameter $ \Lambda_{1S}$
\begin{eqnarray}\label{defnlam1s}
\Lambda_{1S} \equiv \frac{m_{\Upsilon}}{2} - m_b^{\rm 1S}\,.
\end{eqnarray}

The perturbative corrections coming from expressing the $b$ quark mass in terms
of the 1S mass are determined
by using the definitions of $\Lambda_{1S}$ and $m_b^{\rm 1S}$ and the HQET
relationship between meson masses and quark masses
\begin{eqnarray}
m_H = m_{\mathcal{Q}} + \bar{\Lambda} - \frac{\lambda_1 + d_H \lambda_2}{2 \,
m_{\mathcal{Q}}} + \frac{\rho_1 + d_H \rho_2}{4 \, m_{\mathcal{Q}}^2}
- \frac{\tau_1 + \tau_3 + d_H \left(\tau_2 + \tau_4 \right) }{4 \,
m_{\mathcal{Q}}^2}
+ \mathcal{O}\! \left(\frac{\Lambda_{\rm QCD}^4}{m_{\mathcal{Q}}^3} \right),
\end{eqnarray}

\noindent{where $m_H$ is the hadronic mass, $m_{\mathcal{Q}}$ is the heavy
quark mass, and $d_H = 3$ for the pseudoscaler mesons while $d_H = -1$ for the
vector mesons.}

In calculating the general hadronic moment, previously calculated moments by
Bauer,  Ligeti, Luke and Manohar (BLLM) \cite{everyonebutme} were
reexamined in the 1S mass scheme
to check results. The results presented in the following sections for the first
hadronic moment and its variance are different for two reasons.
First, in BLLM a $1/ \bar{m}_B$ expansion was used to replace $m_B$ in the
expansion of $s_H$ in terms of partonic variables,
\begin{eqnarray}
m_B = \bar{m}_B + \frac{3\,\left(m_{\Upsilon} - 4\, \bar{m}_B \right)\,
\lambda_2}{4\,{\bar{m}_B}^2} - \frac{3 \, \lambda_2 \, \Lambda_{1S}}{2\,
{\bar{m}_B}^2}
- \frac{3}{4 \, {\bar{m}_B}^2} \left( \tau_2 + \tau_4 \right) + \frac{3}{4 \,
{\bar{m}_B}^2} \rho_2 +
\mathcal{O}\! \left( \epsilon, \frac{\Lambda_{\rm QCD}^4}{m_{\bar{B}}^3}
\right).
\end{eqnarray}

\noindent{In the results reported in the following sections we always use a $1/
{m_{\Upsilon}}$ expansion. The corresponding expansion is}
\begin{eqnarray}
m_B = \bar{m}_B - \frac{3\, \lambda_2}{{m_{\Upsilon}}} - \frac{6 \, \lambda_2
\, \Lambda_{1S}}{ {m_{\Upsilon}}^2}
- \frac{3}{ {m_{\Upsilon}}^2} \left( \tau_2 + \tau_4 \right) + \frac{3}{
{m_{\Upsilon}}^2} \rho_2
+ \mathcal{O}\! \left( \epsilon, \frac{\Lambda_{\rm QCD}^4}{m_{\Upsilon}^3}
\right).
\end{eqnarray}
Second, we treat a class of powers of 
$(\bar{m}_B - \frac{m_{\Upsilon}}{2})^n$ differently
than  BLLM. When terms are generated by
changing $\bar{\Lambda} \equiv m_B - m_b $ into $\Lambda_{1S}$,
\begin{eqnarray}\label{transeqn}
\bar{\Lambda} = \left(\bar{m}_B  - \frac{m_{\Upsilon}}{2} \right) +
\Lambda_{1S} + \frac{\lambda_1}{m_{\Upsilon}} + \frac{2\, \Lambda_{1S} \,
\lambda_1}{{m_{\Upsilon}}^2}
 + \frac{\left( \tau_1 + \tau_3 - \rho_1 \right)}{{m_{\Upsilon}}^2} +
\mathcal{O}\! \left( \epsilon, \frac{\Lambda_{\rm QCD}^4}{m_{\Upsilon}^3}
\right),
\end{eqnarray}
\noindent{the $(\bar{m}_B - \frac{m_{\Upsilon}}{2})^n$ terms are kept only for
$n \le 3$. In  BLLM this class of terms are treated as $\mathcal{O}\,(1)$
although they are formally of order $\Lambda_{\rm QCD}^n $, leading to  
this class of higher order terms of $(\bar{m}_B - \frac{m_{\Upsilon}}{2})^n$
being kept and summed into the coefficients of the nonperturbative parameters.
In the following results, the $(\bar{m}_B - \frac{m_{\Upsilon}}{2})^n$  terms
from $\bar{\Lambda} $ are counted as order $\Lambda_{\rm QCD}^n$ 
and in the results of Section \ref{moments} they
are  kept only up to $ \mathcal{O}\! \left( \Lambda_{\rm QCD}^3 \right) $ in
the nonperturbative expansion and up to $ \mathcal{O}\! \left( \LQCD \right) $
for perturbative terms. Factors of  $(\bar{m}_B - \frac{m_{\Upsilon}}{2}) $
are also generated in the replacement of the $c$ and $b$ quark mass
and these factors are treated as $\mathcal{O}\,(1)$.
This implementation of the 1S scheme is similar to the 
the general moment analysis of the lepton energy spectrum \cite{christianmike}. 
The lepton moments are presented in Appendix C and can be combined with the
hadronic spectrum results to cross check extractions of $\Lambda_{1S}$ and 
$\lambda_1$ from these differing spectra. Further hadronic moments that are 
appropriate for a global fit, such as $s_H^{1/2}$ and $s_H^{3/2}$ are 
presented in Appendix B.

\subsection{Decay Width to  \orderalphalam,
 $\mathcal{O}\! \left( \Lambda_{\rm QCD}^3  /  {m_b}^3 \right) $ }\label{decay}

With the 1S scheme implemented as discussed in the previous section, the decay
width of \Bdecay  is
\renewcommand{\baselinestretch}{0.75}
{\normalsize{
\begin{eqnarray}
 \Gamma \left( B \to X_c l \bar{\nu} \right)  &=& \Gamma_0
\Big[0.5325 -1.132\, \frac{\Lambda_{1S}}{m_{\Upsilon}/2} - 0.924 \,
\left(\frac{\Lambda_{1S}}{m_{\Upsilon}/2}\right)^2 - 1.89 \,
 \left(\frac{\Lambda_{1S}}{m_{\Upsilon}/2}\right)^3  \nonumber  \\
\,\nonumber \\
&-& 2.12 \,\frac{\lambda_1}{\left( m_{\Upsilon}/2 \right)^2} - 3.93 \,
\frac{\lambda_2}{\left( m_{\Upsilon}/2 \right)^2}
+  0.74 \,\frac{\lambda_1 \, \Lambda_{1S}}{\left( m_{\Upsilon}/2 \right)^3}
- 1.73 \,\frac{\lambda_2 \, \Lambda_{1S}}{\left( m_{\Upsilon}/2 \right)^3}
\nonumber \\
\,\nonumber \\
&-& 5.99 \, \frac{ \rho_1}{\left( m_{\Upsilon}/2 \right)^3}
+ 4.94 \, \frac{ \rho_2}{\left( m_{\Upsilon}/2 \right)^3}  -2.98 \, \frac{
\tau_1}{\left( m_{\Upsilon}/2 \right)^3} + 0.99 \, \frac{ \tau_2}{\left(
m_{\Upsilon}/2 \right)^3}
\, \\
&-& 4.96 \, \frac{ \tau_3}{\left( m_{\Upsilon}/2 \right)^3} -4.94 \, \frac{
\tau_4}{\left( m_{\Upsilon}/2 \right)^3}  -0.080 \, \epsilon
+ 0.133 \, \epsilon \, \frac{\Lambda_{1S}}{m_{\Upsilon}/2}  \nonumber \\
&+& 0.004 \, \epsilon \,  \left( \frac{\Lambda_{1S}}{m_{\Upsilon}/2} \right)^2
\Big] \nonumber ,
\end{eqnarray}
}}
where $\Gamma_0 = \frac{G_{F}^2  |V_{cb}|^2 }{192 \, \pi^3}  \left(
\frac{m_{\Upsilon}}{2} \right)^5$.
Uncertainties in the values of $\Lambda_{1S} $, $\lambda_i$, $\rho_i$ and
$\tau_i$ introduce uncertainties in the
inclusive extractions of $ |V_{cb}| $ using the decay width. In the
nonperturbative expansion the
largest theoretical uncertainty in the extraction of $|V_{cb}|$ comes from
$\Lambda_{1S} $ and $\lambda_1$ which one can see introduce $\sim$ 2$\%$ 
uncertainties, followed by the higher order nonperturbative terms which 
impose an uncertainty of  $\sim$ 1$\%$ as one can see by examining the
results for the total decay width and estimating the size of the unknown 
terms with dimensional analysis.

The size of the  $\mathcal{O}\!\left(\alpha_s^2  \right) $  can be estimated by
calculating the ${\alpha_{s}^2 \, \beta_0}$
contribution to $\mathcal{O}\!\left(\alpha_s^2  \right) $
\cite{twoloopmike,twoloopsmith,everyonebutme},
although these terms are not included in this paper. This has been done
in the 1S scheme for a number of observables and the full size of these
corrections being treated
as an error introduces $ \sim 2\% $ uncertainty in the extraction of $ |V_{cb}|
$.
The next largest uncalculated contributions in the decay width are the
$\mathcal{O}\! \left(\alpha_s \, \lambda_i \right) $ and
$\mathcal{O}\! \left( \alpha_s \, \Lambda_{1S} \right)$ terms and  the
$\mathcal{O}\! \left(\Lambda_{\rm QCD}^4\ / m_b^2 m_c^2 \right)$ neglected
terms.
The size of the \orderalphalamsec terms in the 1S scheme may be estimated by
taking the
size of the $ \alpha_s \, \Lambda_{1S} $ and multiplying by $ \LQCD / m_b  \sim
0.1 $. For the first moment, this indicates
that the order of the $ \alpha_s \, \lambda_i $ terms is expected to be $ \sim
0.01 \,\Lambda_{\rm QCD}^2$
which can be safely neglected in fits to determine the third order parameters
in the OPE. However,
completely uncorrelated uncertainties of this size for both $ \alpha_s \,
\lambda_i $ should be used
to estimate the error on the fit. The size of the corrections introduced when
using the mass splitting
formula to replace the $c$ quark  mass should also be estimated in a fit to
extract the third
order terms in the OPE, as well as uncertainties due to $1/m_b^4$ corrections
to the OPE.
These number and size of these terms are completely unknown and the
uncertainties introduced due to these terms
can be estimated by introducing completely uncorrelated errors of their naive
dimensional size.

\subsection{Integral Hadronic Moments}
The  nonperturbative parameters in the decay width can be determined by global
fits to
moments calculated from the decay spectra of \Bdecay such as the hadronic
invariant mass spectra.
The first and second moments of the hadronic invariant mass spectrum have been
known for some time
\cite{hadronmoment1, hadronmoment3}. The perturbative corrections to these
moments were obtained by expressing the
hadronic moments in terms of the known  \orderalpha $\,$ corrections to the
lepton spectra and the
leading order hadronic invariant mass spectra. This technique fails when a
lepton energy cut is introduced into the
phase space, and the general tensor results of Section \ref{tensor} are
required.
In terms of partonic quantities the hadronic invariant mass is defined to be
\begin{eqnarray}
s_H = \, (Q - q)^2 \,  = {m_B}^2 -  \, m_B \, m_b \, \left( 1 - \hat{z} +
\hat{y} \right) + m_b^2 \, \hat{y} .
\end{eqnarray}

It is conventional to examine the first hadronic moment once the spin averaged
meson mass $\bar{m}_D^2$
is subtracted. Moments that give the mean and  variance of the hadronic
invariant mass spectrum with lepton energy cuts of differing
values were re-examined recently by BLLM. These moments are defined with lepton
energy cuts $\elmin$,
\begin{eqnarray}
 S_1\left( \elmin \right) = \left< s_H - \bar{m}_D^2 \right> \mid_{E_l \ge
\elmin},
\hspace{1cm} S_2\left( \elmin \right) = \left<\left(s_H - \left< s_H \right>
\right)^2 \right> \mid_{E_l \ge \elmin}.
\end{eqnarray}

These moments as functions of the cut on the lepton energy, including the
previously uncalculated perturbative
corrections are as follows. The coefficients stated are for
the dimensionful nonperturbative parameters listed at the top of the column.
The data used in the numerical evaluations  in this paper are $\bar{m}_B =
5.3135 \,\,  \rm{GeV}$,
$\bar{m}_D = 1.9730 \, \, \rm{GeV}$, $m_{\Upsilon} = 9.4603 \, \, \rm{GeV}$
and the strong coupling is $\alpha_s (m_b) = 0.22 $.
For example for the first moment with no cut we find
\begin{eqnarray}
S_1\left( 0 \right) &=& 0.834 + 1.646 \, \Lambda_{1S} + 0.451 \, \Lambda_{1S}^2
+ 0.16 \, \Lambda_{1S}^3 + 1.43 \, \lambda_1 -0.34 \, \lambda_2
+ 0.51 \, \lambda_1 \Lambda_{1S} \nonumber \\
&+&  0.07 \, \lambda_2 \Lambda_{1S} + 0.71 \, \rho_1 - 0.34 \, \rho_2 + 0.32 \,
\tau_1 + 0.25 \, \tau_2 + 0.29 \, \tau_3 + 0.15 \,\tau_4 \nonumber \\
&+& 0.143 \, \epsilon + 0.175 \, \Lambda_{1S} \, \epsilon,
\end{eqnarray}
\noindent{and in the following tables the leading order term of a moment $S_i$
is labelled $S_i^0$.

\begin{center}
\vspace{0.3cm}
\renewcommand{\baselinestretch}{0.75}
{\normalsize{
\begin{tabular}{|c||cccccccccccccc|}
\hline \hline
\, $\elmin$ \, & \, $S_1^0$ \,& \,$\Lambda_{1S}$ \,& \, $\Lambda_{1S}^2$ \,& \,
$\Lambda_{1S}^3$ \,& \, $\lambda_1$ \,& \, $\lambda_2$
 \,&  $\lambda_1 \, \Lambda_{1S}$  &  $\lambda_2 \, \Lambda_{1S}$ & \, $\rho_1$
\,& \, $\rho_2$ \,& \, $\tau_1$
\,& \, $\tau_2$ \,& \, $\tau_3$ \,& \, $\tau_4$ \,  \\
\hline \hline
0 & 0.834 & 1.646 & 0.451 & 0.16 & 1.43 & -0.34 & 0.51 & 0.07 & 0.71 & -0.34 &
0.32 & 0.25 & 0.29 & 0.15 \\
0.5 & 0.822 & 1.623 & 0.445 & 0.16 & 1.44 & -0.30 & 0.51 & 0.09 & 0.72 & -0.34
& 0.32 & 0.26 & 0.29 & 0.16   \\
0.7 & 0.807 & 1.592 & 0.435 & 0.16 & 1.46 & -0.24 & 0.53 & 0.12 & 0.75 & -0.34
& 0.33 & 0.27 & 0.29 & 0.17  \\
0.9 & 0.786 & 1.549 & 0.420 & 0.16 & 1.51 & -0.14 & 0.55 & 0.18 & 0.79 & -0.34
& 0.34 & 0.30 & 0.30 & 0.18  \\
1.1 & 0.762 & 1.496 & 0.397 & 0.15 & 1.57 & 0.00 & 0.59 & 0.26 & 0.87 & -0.33 &
0.35 & 0.34 & 0.31 & 0.21   \\
1.3 & 0.737 & 1.439 & 0.368 & 0.15 & 1.69 & 0.18 & 0.66 & 0.37 & 0.99 & -0.30 &
0.37 & 0.41 & 0.32 & 0.24   \\
1.5 & 0.719 & 1.392 & 0.334 & 0.14 & 1.92 & 0.42 & 0.79 & 0.51 & 1.23 & -0.23 &
0.41 & 0.53 & 0.33 & 0.28 \\
\hline
\end{tabular}
}} \\
\vspace{0.5cm}
TABLE 1: Coefficients of the nonperturbative parameters for $S_1 \left(E_0
\right)$. \\
\vspace{1cm}
\renewcommand{\baselinestretch}{0.75}
{\normalsize{
\begin{tabular}{|c||cc|cc|cc|}
\hline \hline &
\multicolumn{2}{|c|}{ \rm{1S ${\alpha_s}^2$ Contribution}} &
\multicolumn{2}{|c|}{ \rm{ ${\alpha_s}$ Contribution} } &
\multicolumn{2}{|c|}{ \rm{Combined $O \left(\epsilon \right)$}}
\\ \hline
\, $\elmin$ \, & \,  $\epsilon$ \, & \,$\Lambda_{1S}$  $\epsilon $ \, &
 \,  $ \epsilon$ \, & \,$\Lambda_{1S}$  $\epsilon $ \, &  \,   $\epsilon$ \, &
\, $\Lambda_{1S}$  $\epsilon $ \, \\
\hline \hline
0 & -0.014 & 0.090 &  0.157 & 0.086 & 0.143  &  0.175 \\
0.5 & -0.014 & 0.088 & 0.143 & 0.069 & 0.129  & 0.157  \\
0.7 & -0.014 & 0.086 & 0.139 & 0.072 & 0.125  &  0.159  \\
0.9 & -0.014 & 0.084 & 0.134 &  0.076 & 0.120  &  0.160  \\
1.1 & -0.014 & 0.081 & 0.128 & 0.080 & 0.114  & 0.161  \\
1.3 & -0.015 & 0.077 & 0.121 & 0.085 & 0.106  & 0.162  \\
1.5 & -0.018 & 0.073 & 0.117 & 0.093 & 0.099  & 0.166  \\
\hline
\end{tabular}
}} \\
\vspace{0.5cm}
TABLE 2: Coefficients of the perturbative parameters for $S_1 \left(E_0
\right)$. \\
\end{center}

\vspace{0.5cm}
For the $\rm{S_2 \left( E_0 \right)}$ moments, as explained in Section
\ref{upsilon}, the results differ from
those stated in BLLM. This difference is formally of higher order, and in the
nonperturbative expansion
the overall effect of the differing implementations of the 1S scheme is small.
The effect of these terms in the perturbative expansion is also small for most
moments.
However, the variance of the hadronic invariant mass spectrum is more sensitive
to higher order terms
due to the cancellation among leading order terms in the nonperturbative
expansion.
The 1S scheme as implemented in BLLM found that the variance increased as the
lepton energy cut was increased
due to the dominance of the \orderalpha $\,$ term and the suppression of
leading order nonperturbative corrections.
The  \orderalpha $\,$ term is the dominant correction to the variance in the
$m_b \rightarrow \infty$ limit.
The experimentally measured lepton energy cut dependence has the variance
decreasing as the lepton energy cut increases.
When  $(\bar{m}_B - \frac{m_{\Upsilon}}{2})$ terms are treated as detailed in 
Section \ref{upsilon} in the 1S scheme, the \orderalpha $\,$ term and variance 
has the experimentally measured dependence on the lepton energy cut, as can be
seen in Table 3 and 4.
\begin{center}
\vspace{0.5cm}
\renewcommand{\baselinestretch}{0.75}
{\normalsize{
\begin{tabular}{|c||cccccccccccccc|}
\hline \hline
 $\elmin$  & \, $S_2^0$ \,& \,$\Lambda_{1S}$ \,& \, $\Lambda_{1S}^2$ \,& \,
$\Lambda_{1S}^3$ \,& \, $\lambda_1$ \,& \, $\lambda_2$ \,&  $\lambda_1 \,
\Lambda_{1S}$
&  $\lambda_2 \, \Lambda_{1S}$ & \, $\rho_1$ \,& \, $\rho_2$ \,& \, $\tau_1$
\,& \, $\tau_2$ \,& \, $\tau_3$ \,& \, $\tau_4$ \,  \\
\hline \hline
0   & 0.0163 & 0.09 & 0.08 & -0.03 & -4.87 & 1.33 & -1.85 & 2.04 & -6.41 & 1.51
& -1.04 & -2.87 & 0.00 & 0.25 \\
0.5 & 0.0152 & 0.08 & 0.07 & -0.03 & -4.78 & 1.34 & -1.79 & 2.07 & -6.48 & 1.39
& -1.03 & -2.82 & 0.00 & 0.25 \\
0.7 & 0.0153 & 0.08 & 0.07 & -0.03 & -4.65 & 1.35 & -1.70 & 2.09 & -6.61 & 1.23
& -1.01 & -2.77 & 0.00 & 0.25 \\
0.9 & 0.0167 & 0.09 & 0.07 & -0.04 & -4.48 & 1.37 & -1.56 & 2.12 & -6.84 & 0.99
& -0.98 & -2.69 & 0.00 & 0.25 \\
1.1 & 0.0195 & 0.10 & 0.08 & -0.04 & -4.26 & 1.40 & -1.36 & 2.17 & -7.23 & 0.68
& -0.95 & -2.61 & 0.00 & 0.24 \\
1.3 & 0.0224 & 0.11 & 0.09 & -0.04 & -4.01 & 1.44 & -1.10 & 2.25 & -7.88 & 0.28
& -0.92 & -2.53 & 0.00 & 0.24 \\
1.5 & 0.0227 & 0.11 & 0.09 & -0.04 & -3.76 & 1.52 & -0.77 & 2.38 & -9.05 &
-0.22 & -0.91 & -2.48 & 0.00 & 0.24 \\
\hline
\end{tabular}
}} \\
\vspace{0.5cm}
 TABLE 3: Coefficients of the nonperturbative parameters for $S_2 \left(E_0
\right)$. \\

\vspace{1cm}
\renewcommand{\baselinestretch}{0.75}
{\normalsize{
\begin{tabular}{|c||cc|cc|cc|}
\hline \hline &
\multicolumn{2}{|c|}{ \rm{1S ${\alpha_s}^2$ Contribution}} &
\multicolumn{2}{|c|}{ \rm{ ${\alpha_s}$ Contribution} } &
\multicolumn{2}{|c|}{ \rm{Combined $O \left(\epsilon \right)$}}
\\ \hline
\, $\elmin$ \, & \,  $\epsilon$ \, & \,$\Lambda_{1S}$  $\epsilon $ \, &
 \,  $ \epsilon$ \, & \,$\Lambda_{1S}$  $\epsilon $ \, &  \,   $\epsilon$ \, &
\, $\Lambda_{1S}$  $\epsilon $ \, \\
\hline \hline
0 & 0.105 & 0.163 & 0.551 & 0.424 & 0.656  & 0.588  \\
0.5 & 0.102 & 0.159 &  0.163 & -0.202 & 0.265  & -0.043  \\
0.7 & 0.099 & 0.154 &  0.099 & -0.266 & 0.198  & -0.112  \\
0.9 & 0.094 & 0.147 &  0.057 & -0.283 & 0.151  & -0.136  \\
1.1 & 0.089 & 0.140 &  0.029 & -0.273 & 0.118  & -0.132  \\
1.3 & 0.085 & 0.135 &  0.010 & -0.240 & 0.096  & -0.112  \\
1.5 & 0.083 & 0.133 &  0.002 & -0.211 & 0.083  & -0.078  \\
\hline
\end{tabular}
}}\\
\vspace{0.5cm}
TABLE 4: Coefficients of the perturbative parameters for $S_2 \left(E_0
\right)$. \\
\end{center}
The moments $S_1$ and $S_2$ with a lepton energy cut of  $1.5 \, \rm{GeV}$ have
been experimentally measured by CLEO \cite{CLEO20012},
\begin{eqnarray}
S_1 \left( 1.5 \, {\rm GeV } \right) = 0.251 \, \pm \, 0.066 \,  \rm{GeV}^2 ,
\\
S_2 \left( 1.5 \, {\rm GeV } \right) = 0.576 \, \pm \, 0.170  \, \rm{GeV}^4 ,
\end{eqnarray}

\noindent{Using this data, we can extract values of $\Lambda_{1S}$ and
$\lambda_1$,
for comparison with extractions from the lepton energy spectrum. We estimate
the theoretical error on the extraction due to the unknown
third order terms in the usual way \cite{hadron3rdorder} using the results of
recent fits when they substantially
improve our knowledge of these terms beyond dimensional analysis.
We use the HQET vector pseudoscalar mass splitting constraint to determine
$\lambda_2 = 0.12 \, {\rm{GeV^2}} $
and the  mass splitting formula to third order \cite{hadronmoment3} ,}

\begin{eqnarray}
\rho_2 - \tau_2 - \tau_4 &=&
\frac{\kappa\left(m_c \right)\,  m_b^2 \, m_c \, \Delta m_B - m_b \, m_c^2 \,
 \Delta m_D}{m_b  - m_c \, \kappa\left(m_c \right)},
\end{eqnarray}
\noindent{to reduce the number of free parameters. A positive value of $\rho_1$
is imposed in accordance
with vacuum saturation \cite{vacuumsat} and is drawn from the range $\left[0,
0.125 \right]  \, \rm{GeV^3} $.
The unknown matrix elements are then drawn from a flat distribution,
the unknown third order terms are drawn from between $\pm 0.125 \, {\rm
GeV}^3 $ while $\lambda_1$ in drawn from  $\left[-600, 0 \right]  \, \rm{{\rm
MeV}^2} $
in accordance with its full constrained range from the BLLM fit. We then
extract the following values for $ \lambda_1 $ and $ \Lambda_{1S} $:
\begin{eqnarray}
\Lambda_{1S} \!\!\!&=&\!\!\! \left[-0.13 \,  \pm 0.05_\epsilon \pm
0.09_{m^3}\right] {\rm GeV}\nn\\
\lambda_1 \!\!\!&=&\!\!\! \left[-0.24 \,   \pm 0.02_\epsilon \pm
0.09_{m^3}\right] {\rm GeV}^2\,.
\end{eqnarray}

The perturbative errors are estimated by using the two loop running of
$\alpha_s $ to vary the scale of $\alpha_s \left( \mu \right)$ between $m_b/2 <
\mu < 2m_b$.
Adding the theoretical errors in quadrature we obtain $m_b^{\rm 1S} = 4.86 \pm
0.10 \, {\rm GeV}$, which is in excellent agreement with the inclusive
extraction using the lepton energy moments, $m_b^{\rm 1S} = 4.84 \pm 0.10 \,
{\rm GeV}$  \cite{christianmike}  and in agreement with the results of BLLM where
$m_b^{\rm 1S} = 4.74 \pm 0.10 \, {\rm GeV}$, despite the differences in the
expansion and the larger number of observables in the fit. This extracted
1S mass $ m_b^{\rm 1S} = 4.86 \pm 0.10 \, {\rm GeV} $ translates into a value of  
the $ {\rm \overline{MS} } $ mass $ \bar{m}_b(m_b) =  4.34 \pm 0.09 \, {\rm GeV} $ 
which  can be compared with other extractions of the $ {\rm \overline{MS} } $ mass 
\cite{lukebmass} such as  the $ {\rm \overline{MS} } $ mass
found by examining moments of the $b \bar{b} $ production cross section 
\cite{valoshinbb,RussianMoments,Pineda, leutwyler, BrambillaMoments}
a recent analysis of which finds  $ \bar{m}_b(m_b) =  4.25 \pm 0.08 \, {\rm GeV} $ 
\cite{benekesinger}.

\subsection{Fractional Moments}
\subsubsection{ The $1/m_c $ Expansion}

In integer moments such as $S_1$ and $S_2$, a $\LQCD/m_c$ expansion only enters
the predicted value of a moment through the mass transformation relationship
Eq.  ( \ref{charmrelation} ). Fractional moments have additional cuts in the
complex
$q \cdot v$ plane due to the branch cut starting at $s_H = 0$ when $s_H$ is
taken to a noninteger power.
These branch cuts are separated from the physical cut by a scale set by $m_c$,
as the physical branch cut begins at $s_H = m_D^2$.
As $m_c \rightarrow 0$ these cuts coalesce and one would expect predictions for
fractional moments in this limit to be ill defined,
as discussed in recent work \cite{uraltsev}. We find an explicit 
$m_b \, \LQCD/m_c^2$
expansion in calculations of fractional moments of $s_H$; the neglected terms 
in this expansion are numerically suppressed for hadronic
invariant mass observables leading to a small numerical uncertainty being
introduced.
This can be shown by examining a general moment of the squared hadronic
invariant mass $s_H^n$.
The dependence of the general moment as a function of $n$ is found by
performing an expansion of $s_H^n$,
\begin{eqnarray}
s_H^n =  \sum_{k = 0}^{\infty} \sum_{l = 0}^k \, \frac{\Gamma(n+ 1)}{ \Gamma(n+
1 - k)  \, \Gamma(k)} \, C^k_l \, \hat{y}^l \, \hat{z}^{n-k} \, m_b^{2 n} .
\end{eqnarray}
The coefficient functions $ C^k_l $ are $ O \left( \Lambda_{\rm QCD}^k /m^{k}_b
\right)$. Expanding up to $ O \left( \Lambda_{\rm QCD}^3 / m_b^3 \right)$
in the nonperturbative expansion we find,

\begin{eqnarray}
s_H^n = \,\hat{z}^n \, m_b^{(2 \, n)} \,\Big[ C_0^0 + \frac{n}{\hat{z}} \left(
C_0^1 + \hat{y} \, C_1^1 \right)
+ \frac{n \, \left( n-1 \right)}{1! \,\hat{z}^2} \left(C_0^2 + \hat{y} \, C_1^2
+ \hat{y}^2 \, C_2^2 \right)  \nonumber \\
 + \frac{n \, \left( n-1 \right) \,  \left( n-2 \right)}{ 2! \,\hat{z}^3}
\left(C_0^3 + \hat{y} \, C_1^3 + \hat{y}^2 \, C_2^3 + \hat{y}^3 \, C_3^3
\right) \Big],
\end{eqnarray}
where the $C_l^k$ are functions of $n$  and the nonperturbative matrix elements
\footnote{These $C_i^i$  coefficient functions are reported in the Appendix.}.
For integer moments this expression has no $1/z$
dependence. However, for non-integer moments in this range one obtains
contributions
of order $ z^{-k} $ where $k \ge n$ is the ceiling of the fractional moment
power $n$. As the lower limit of $z$
is $\rho = m_c^2  / {m_b^2}$,  this corresponds to a $m_b \, \LQCD / m_c$ 
expansion entering into the calculations of fractional
moments. This  expansion does not seem to introduce a large uncertainty
for fractional moments compared to integer moments
as the neglected class of terms are numerically suppressed 
for $n$ values in the range $[0,3]$ but the best way to estimate the uncertainty 
introduced is under study. In the following investigation of hadronic 
fractional moments no additional uncertainty is added to account for this 
theoretical error and we examine how known sources of error primarily from unknown 
matrix elements can be reduced.

When examining a general moment $s_H^n$ to obtain interesting observables, we
expand in the ratio $\LQCD/ m_{\mathcal{Q}}$  and then examine the
$n$ and $\elmin$ dependence of the coefficient functions of the nonperturbative
matrix elements. The essential observation
motivating this approach is that one is allowed to choose $n$ and $\elmin$
within a range of reasonably accessible
experimental values, in order to maximize the utility of a measured moment in
obtaining information on the
nonperturbative matrix elements. We define a general moment function,
\begin{eqnarray} \label{momentdefinition}
S [n,E_{\ell_1},m,E_{\ell_2}] = \frac{\left< s_H^n \right> \mid_{E_l \ge
E_{1}}}{\left< s_H^m \right> \mid_{E_l \ge E_{2}}} ,
\end{eqnarray}
\noindent{so that}
\begin{eqnarray}
 S_1\left( \elmin \right) &=& S [1,\elmin,0,\elmin] - \bar{m}_D^2 \nonumber, \\
 S_2\left( \elmin \right) &=& S [2,\elmin,0,\elmin] - \left(S
[1,\elmin,0,\elmin] \right)^2. \nonumber
\end{eqnarray}
Our search of the hadronic mass moments is restricted to the parameter space,
\begin{eqnarray}
{m}<3\,,\, n<3\,, \quad 0.5\, {\rm GeV} < E_{\ell_i}^{\rm cut}< 1.5\, {\rm
GeV},\,
\end{eqnarray}
\noindent{ to ensure a well behaved OPE. In this parameter space we find  two
types of
moments of interest, moments that allow the OPE to be precisely tested for
deviations from experiment and moments that
allow one to extract the 1S mass with minimal error. We consider each type of
moment in the following sections. }

\subsubsection{OPE Testing Moments}

A discrepancy of the prediction of the OPE when compared with data can come
from a number of possible sources
when one is considering percent level extractions of $|V_{cb}|$:
a higher order matrix element that is being neglected could be anomalously
large, the OPE
itself could not be converging or quark-hadron duality violation could effect
determinations \cite{Yaouanc1,Yaouanc2,Yaouanc3,Bigi,Bigi2,Isgur}.
By finding moments where only the leading order unknown nonperturbative
parameters are suppressed and checking the predicted values against experiment,
one can assess the theoretical error that is being assigned to the inclusive
extraction of $|V_{cb}|$ in a clear and unambiguous fashion.
This technique \cite{christianmike} has recently been used to test the OPE in
the lepton
energy spectrum. Measurements of these OPE testing observables in this spectrum
indicated that
the OPE is a valid description of the data to the percent level
\cite{CLEO2002m}.
It is important to note that the OPE testing moments presented allow one to
check
that the error assigned in the extractions of $|V_{cb}|$ and $m_b$ is large
enough to account for all
of these possible effects if the OPE testing moments properly describe the data
\cite{artuso}.
\newpage
A selection of these OPE testing moments for the hadronic invariant mass
spectrum is presented in Table 5. The nonperturbative parameters for these 
moments are not suppressed except for
the leading unknown nonperturbative terms
$\Lambda_{1S}$ and  $\lambda_1 $. A typical OPE testing moment is as follows,
\renewcommand{\baselinestretch}{0.75}
\begin{eqnarray}
D_a^1 &=& S[2,0.5,2.2,0.7] \nonumber \\
&=& \, 0.7779
\Big[ 1 + 0.168\, \frac{\Lambda_{1S}}{m_{\Upsilon}/2} + 0.345 \,
\left(\frac{\Lambda_{1S}}{m_{\Upsilon}/2}\right)^2 + 0.501 \,
 \left(\frac{\Lambda_{1S}}{m_{\Upsilon}/2}\right)^3  +0.01
\, \frac{\lambda_1}{\left( m_{\Upsilon}/2 \right)^2}  \nonumber  \\
&\,& \hspace{2cm} - 1.05 \, \frac{\lambda_2}{\left( m_{\Upsilon}/2 \right)^2} -
 1.94 \,\frac{\lambda_1 \, \Lambda_{1S}}{\left( m_{\Upsilon}/2 \right)^3}
- 8.75 \,\frac{\lambda_2 \, \Lambda_{1S}}{\left( m_{\Upsilon}/2 \right)^3}
+6.18 \, \frac{ \rho_1}{\left( m_{\Upsilon}/2 \right)^3} \nonumber \\
\,\nonumber \\
&\,& \hspace{2cm} - 0.25 \, \frac{ \rho_2}{\left( m_{\Upsilon}/2 \right)^3} +
0.27 \, \frac{ \tau_1}{\left( m_{\Upsilon}/2 \right)^3}
+ 3.71 \, \frac{ \tau_2}{\left( m_{\Upsilon}/2 \right)^3}
-1.50 \, \frac{ \tau_3}{\left( m_{\Upsilon}/2 \right)^3}  \nonumber \\
\, \nonumber \\
&\,& \hspace{2cm} -1.61 \, \frac{ \tau_4}{\left( m_{\Upsilon}/2 \right)^3}  -
0.0094 \, \epsilon   - 0.008 \, \epsilon \, \frac{\Lambda_{1S}}{m_{\Upsilon}/2}
 \Big].
\end{eqnarray}

Varying the unknown parameters in the same way, and treating the
leading order nonperturbative parameters as unknowns and varying them
over the region $\Lambda_{1S} = -0.13 \pm 0.1 \, {\rm GeV}$, $\lambda_1  = -0.3
\pm 0.3 \,{\rm GeV^2}$, this moment is predicted to be
}}

\begin{eqnarray}
D_a^1 &\!\!=\!\!& 0.7686  \pm 0.0018_\epsilon \pm 0.0040_{{\rm N.P.}}  \,.
\end{eqnarray}
\noindent{ The perturbative error is obtained by the scale variation in the
standard way.}
As in the lepton spectra, the OPE testing moments are such that with no
nonperturbative input other than the known
value of $\lambda_2$, we can predict the value of a moment to an accuracy of
1$\% $.

As the error on the nonperturbative terms is reduced with global fits, it is
important to cross check with the
predicted and measured values of these moments in order to ensure the
error on $|V_{cb}| $ and $m_b^{ \rm 1S}$ is not being underestimated.  Table 5
and 6 present a selection of
OPE testing moments in the parameter space examined.

\begin{center}
\vspace{0.5cm}
\renewcommand{\baselinestretch}{0.75}
{\normalsize{
\begin{tabular}{|c||cccccccccccccc|}
\hline \hline
 \, & \,  \,& \,$\Lambda_{1S}$ \,& \, $\Lambda_{1S}^2$ \,& \, $\Lambda_{1S}^3$
\,& \, $\lambda_1$ \,& \,
$\lambda_2$ \,&  $\lambda_1 \, \Lambda_{1S}$ &  $\lambda_2 \, \Lambda_{1S}$ &
\, $\rho_1$ \,& \, $\rho_2$ \,& \, $\tau_1$ \,& \,
$\tau_2$ \,& \, $\tau_3$ \,& \, $\tau_4$ \,  \\
\hline \hline
$D_a^1 $ & 0.7779 & -0.028 & 0.01 & 0.00 & 0.00 & -0.04 & -0.01 & -0.06 & 0.05
& -0.00 & 0.00 & 0.03 & -0.01 & -0.01 \\
$D_b^2 $ & 0.8845 & -0.019 & 0.00 & 0.00 & 0.00 & -0.02 & -0.01 & -0.03 & 0.03
& 0.00 & 0.00 & 0.01 & -0.01 & -0.01 \\
$D_c^3 $ & 0.7829 & 0.047 & 0.10 & 0.04 & 0.00 & -0.13 & -0.03 & -0.16 & 0.06 &
0.03 & 0.00 & 0.03 & -0.02 & -0.02 \\
$D_d^4 $ & 1.9030 & 0.166 & -0.08 & -0.03 & 0.00 & 0.22 & 0.08 & 0.38 & -0.27 &
0.00 & -0.01 & -0.15 & 0.06 & 0.07 \\
$D_e^5 $ & 2.428 & 0.040 & -0.46 & -0.20 & 0.00 & 0.66 & 0.19 & 0.99 & -0.62 &
-0.08 & -0.03 & -0.29 & 0.12 & 0.14 \\
\hline
\end{tabular}
}} 
\vspace{0.25cm}
\end{center}
TABLE 5: Coefficients of the nonperturbative parameters for OPE testing
Moments. The OPE testing moments in the
table are defined in the following way, $D_a^1  = S\left[2.0,0.5,2.2,0.7
\right]$,
$D_b^2 = S\left[1.9,0.6, 2, 0.7 \right]$ , $D_c^3  = S\left[2.6,0.6, 2.9, 1
\right]$,
$D_d^4  = S\left[2.4,1, 1.9, 0.8 \right]$  and $D_e^5  = S\left[2.9,1.4, 2.2,
1.3 \right]$.
\begin{center}
\vspace{0.75cm}
\renewcommand{\baselinestretch}{0.75}
{\normalsize{
\begin{tabular}{|c||cc|cc|cc|c|}
\hline \hline &
\multicolumn{2}{|c|}{ \rm{1S ${\alpha_s}^2$ Term }} &
\multicolumn{2}{|c|}{ \rm{ ${\alpha_s}$ Term} } &
\multicolumn{2}{|c|}{ \rm{Combined $O \left(\epsilon \right)$}} &
\rm{Predicted Value for Moment}
\\ \hline
\, Label & \,  $\epsilon$ \, & \,$\Lambda_{1S}$  $\epsilon $ \, &
 \,  $ \epsilon$ \, & \,$\Lambda_{1S}$  $\epsilon $ \, &  \,   $\epsilon$ \, &
\, $\Lambda_{1S}$  $\epsilon $ \,
&  \, \\
\hline \hline
$D_a^1 $ & -0.001 & -0.005 & -0.006 & -0.001 & -0.007 & -0.007 &  0.7686 $\pm
0.0018 \, \epsilon \, \pm 0.0040 ({\rm N.P.}) $  \\
$D_b^2 $ & -0.001 & -0.003 & -0.005 & -0.004 & -0.006 & -0.007 & 0.8804 $\pm
0.0014 \, \epsilon \, \pm 0.0025 ({\rm N.P.}) $ \\
$D_c^3 $ & -0.001 & -0.004 & -0.005 & -0.012 & -0.006 & -0.016 & 0.7582 $\pm
0.0016 \, \epsilon \, \pm 0.0067 ({\rm N.P.}) $  \\
$D_d^4 $ & 0.006 & 0.029 & 0.025 & -0.012 & 0.030 & 0.017 & 1.9448 $\pm 0.0078
\, \epsilon \, \pm 0.026 ({\rm N.P.}) $   \\
$D_e^5 $ & 0.010 & 0.050 & 0.026 & -0.036 & 0.036 & 0.014 & 2.5380 $\pm 0.0102
\, \epsilon \, \pm 0.047 ({\rm N.P.}) $ \\
\hline
\end{tabular}
}}\\
\vspace{0.25cm}
\end{center}
TABLE 6: Coefficients of the perturbative parameters for OPE testing moments
and their predicted value.

\subsubsection{Moments to measure $m_b^{\rm 1S}$ with minimal error }
Moments that allow a direct measurement of the $b$ quark mass with minimal
dependence on the unknown $\lambda_1$
nonperturbative parameter have also been found. Moments of this type are
important as a
measurement of $m_b^{\rm 1S}$ to this precision will be an important step in
reducing
the error on $|V_{ub}|$ as well as extracting $|V_{cb}|$.
These moments are particularly suited to being used in a fit to
extract the 1S mass. An example of this type of moment is

\renewcommand{\baselinestretch}{0.75}
{\normalsize{
\begin{eqnarray}
B_a^1 &=& S[3,0.5,0.5,0.9] \nonumber \\
&=& \hspace{0.1cm} 51.9318 \Big[ 1 + 3.215\,
\frac{\Lambda_{1S}}{m_{\Upsilon}/2} + 2.609 \,
\left(\frac{\Lambda_{1S}}{m_{\Upsilon}/2}\right)^2 - 2.216 \,
 \left(\frac{\Lambda_{1S}}{m_{\Upsilon}/2}\right)^3
+0.40\,\frac{\lambda_1}{\left( m_{\Upsilon}/2 \right)^2} \nonumber  \\
\,\nonumber \\
&+& 3.67 \, \frac{\lambda_2}{\left( m_{\Upsilon}/2 \right)^2} +  6.14
\,\frac{\lambda_1 \, \Lambda_{1S}}{\left( m_{\Upsilon}/2 \right)^3}
+ 45.49 \,\frac{\lambda_2 \, \Lambda_{1S}}{\left( m_{\Upsilon}/2 \right)^3}
-27.78 \, \frac{ \rho_1}{\left( m_{\Upsilon}/2 \right)^3}
+ 4.10 \, \frac{ \rho_2}{\left( m_{\Upsilon}/2 \right)^3} \nonumber \\
\,\nonumber \\
&-&  0.004 \, \frac{ \tau_1}{\left( m_{\Upsilon}/2 \right)^3} - 29.15 \, \frac{
\tau_2}{\left( m_{\Upsilon}/2 \right)^3}
+13.26 \, \frac{ \tau_3}{\left( m_{\Upsilon}/2 \right)^3} +10.64 \, \frac{
\tau_4}{\left( m_{\Upsilon}/2 \right)^3} \nonumber \\
\, \nonumber \\
&+& 0.084 \, \epsilon  + 0.154 \, \epsilon \,
\frac{\Lambda_{1S}}{m_{\Upsilon}/2}  \Big] .
\end{eqnarray}
}}
Note that moments of this type have a strong dependence on 
$\Lambda_{1S} $ and  a weak dependence on $\lambda_1$
while the coefficients of  the higher order terms in the 
nonperturbative series are not suppressed.
These moments are thus well suited to measure the b quark 
mass with minimal error as the largest source of theoretical 
error is suppressed in a controlled fashion.
Estimating the error on the extraction in the usual way one finds  
the $m_b^{\rm 1S}$ mass extracted from  this moment
will have a theoretical error due to unknown matrix elements and 
perturbative terms of 
$\pm 50 \, {\rm MeV} ({\rm N.P.}) \,  \pm 3 \, {\rm MeV} \, (\epsilon) $, 
where the  error is dominated by the unknown nonperturbative 
corrections at third order. 
Adding these errors in quadrature one obtains a theoretical error in
the extraction $\sim 50 \, {\rm MeV } $. It is important to experimentally
measure moments of this type for a precise value of the $b$ quark mass to 
be extracted from this spectrum. This error assessment in the extraction of 
$ m_b^{\rm 1S} $ is assigned in accordance with how
theoretical error is assigned in the OPE testing moments.
By measuring both the OPE testing moments and the moments presented in 
this section, one can extract $ m_b^{\rm 1S} $ with an experimentally 
tested theoretical error;  comparisons of the $b$ quark mass extracted 
in this way with  extractions using other techniques will be a useful 
cross check of theoretical techniques.
A selection of moments of this type is given in Table 7 and 8.
\begin{center}
\renewcommand{\baselinestretch}{0.75}
{\normalsize{
\begin{tabular}{|c||cccccccccccccc|}
\hline \hline
 & \,  \,& \,$\Lambda_{1S}$ \,& \, $\Lambda_{1S}^2$ \,& \, $\Lambda_{1S}^3$ \,&
\, $\lambda_1$ \,& \,
$\lambda_2$ \,&  $\lambda_1 \, \Lambda_{1S}$ &  $\lambda_2 \, \Lambda_{1S}$ &
\, $\rho_1$ \,& \, $\rho_2$ \,& \, $\tau_1$ \,& \,
$\tau_2$ \,& \, $\tau_3$ \,& \, $\tau_4$ \,  \\
\hline \hline
$B_a^1 $  & 51.9318 & 35.300 & 6.06 & -1.09 & 0.93 & 8.51 & 3.01 & 22.32 &
-13.63 & 2.01 & 0.00 & -14.30 & 6.51 & 5.22 \\
$B_b^2 $  & 10.2470 & 4.850 & 1.38 & 0.37 & 0.02 & 1.74 & 0.43 & 3.25 & -2.84 &
-0.03 & -0.04 & -2.03 & 0.88 & 0.72 \\
$B_c^3 $  & 12.2674 & 6.116 & 1.48 & 0.28 & 0.31 & 1.35 & 0.79 & 3.62 & -3.50 &
0.24 & 0.00 & -2.43 & 1.10 & 0.88 \\
$B_d^4 $  & 4.7852 & 1.659 & 0.55 & 0.20 & 0.10 & 0.55 & 0.16 & 0.93 & -0.95 &
-0.11 & 0.01 & -0.62 & 0.30 & 0.23 \\
\hline
\end{tabular}
}} \\
\end{center}
TABLE 7 : The nonperturbative parameters of moments to measure the 1S mass
accurately.
 In the table above the moments are defined in the following way $B_a^1  =
S\left[3,0.5,0.5,0.9 \right]$, $B_b^2  = S\left[2.4,0.5,1.2,1.3 \right]$,
$B_c^3  = S\left[2.5,0.5,1,1 \right]$ and $B_d^4  = S\left[2, 0.5,1.3,1.3
\right]$.\\
\vspace{.5cm}
\begin{center}
\renewcommand{\baselinestretch}{0.75}
{\normalsize{
\begin{tabular}{|c||cc|cc|cc|c|}
\hline \hline &
\multicolumn{2}{|c|}{ \rm{1S ${\alpha_s}^2$ Term}} &
\multicolumn{2}{|c|}{ \rm{ ${\alpha_s}$ Term} } &
\multicolumn{2}{|c|}{ \rm{Combined $O \left(\epsilon \right)$}} &
\rm{Error in extraction of 1S mass }
\\ \hline
 & \,  $\epsilon$ \, & \,$\Lambda_{1S}$  $\epsilon $ \, &
 \,  $ \epsilon$ \, & \,$\Lambda_{1S}$  $\epsilon $ \, &  \,   $\epsilon$ \, &
\, $\Lambda_{1S}$  $\epsilon $ \,
&  \, \\
\hline \hline
$B_a^1 $ & -0.003 & 2.231 & 4.345 & -0.536  & 4.343 & 1.70 & $\pm 3 \, {\rm
{\rm MeV}}  \, \epsilon \, \pm 50 \, {\rm MeV} \, ({\rm N.P.}) $  \\
$B_b^2 $ & 0.042 & 0.353 & 0.383 & -0.336 & 0.426 & 0.017 & $\pm 5 \, {\rm MeV}
 \, \epsilon \, \pm 55 \, {\rm MeV} \, ({\rm N.P.}) $ \\
$B_c^3 $ & 0.042 & 0.454 &  0.564 & -0.318 & 0.606 & 0.137 & $\pm 1 \, {\rm
MeV}  \, \epsilon \, \pm 54 \, {\rm MeV} \, ({\rm N.P.})  $  \\
$B_d^4 $ & 0.017 & 0.121 & 0.090 & -0.158 & 0.107 & -0.004 & $\pm 2 \, {\rm
MeV}  \, \epsilon \, \pm 59 \, {\rm MeV} \, ({\rm N.P.})  $   \\
\hline
\end{tabular}
}} \\
\vspace{0.75cm}
\end{center}
TABLE 8: The perturbative parameters of moments to measure the 1S mass
accurately, and the error estimated in using the moment
to extract the 1S mass.\\
\vspace{0.5cm}

\section{Conclusions}

We have presented the  \orderalpha $ \,$ corrections to the structure functions
of the
hadronic tensor for \Bdecay.
The \orderalpha $\,$ and \orderalphalam $\,$perturbative corrections for
lepton energy moments and hadronic invariant mass moments have been calculated.
The effects of the charm quark expansion in fractional moments was shown to be
small and moments that
allow one to extract the nonperturbative parameters revelant for a percent
level determination of
$|V_{cb}|$ from inclusive $B$ decay were presented.
Using the techniques outlined, a $b$ quark mass measurement with a theoretical
error at the $50 \,{\rm MeV} $ or better should
be possible using the inclusive semileptonic decay data. This theoretical
error
assessment, including the assumption of negligible quark-hadron duality
violation  that this analysis relies upon,
is directly testable with the  OPE testing moments presented in this paper.
Fits based on the
results presented should allow an extraction of $|V_{cb}|$ with $\sim \,2 \% $
theoretical error.
As the lepton energy cut dependence of the \orderalphalam term is now known,
the
largest estimated theoretical uncertainty in inclusive extractions of
$|V_{cb}|$ and $m_b^{\rm 1S}$
comes from the $\mathcal{O}\!\left(\alpha_s^2 \right)$ corrections.
\newpage
\begin{center}
\bf{ACKNOWLEDGMENTS}
\end{center}

 It is a pleasure to acknowledge Christian Bauer for collaboration in the early
stages of this work,
for pointing out the importance of the fractional moment charm quark issues and
for comments on the manuscript.
I am also grateful to Michael Luke, Craig Burrell and particularly Alex Williamson 
for helpful discussions and comments on the manuscript.

This work was supported in part by the Walter B. Sumner foundation.

\appendix
\section{$s_H$ Expansion}
The coefficient functions of the general hadronic moment $s_H$ in terms of the
pole mass
and the HQET local operator expansion are as follows:

\begin{eqnarray}
C_0^0 = 1 &+& \frac{n \, \Lambda}{{m_b}} - \frac{n \,\left( \lambda_1 + 3 \,
\lambda_2  - ( n - 1) \Lambda^2 \right)}{2 \, {m_b}^2} \
+ {\frac{\left(n -1  \right) \,n\,\Lambda \,\left( \left( n -2  \right)
\,{{\Lambda }^2}
- 3\,\left( \lambda 1 + 3\, \lambda 2 \right)  \right) }{6\,{{m_b}^3}}}
\nonumber \\
&-&\frac{  n\,\left( \tau_1 + 3\,\tau_2 + \tau_3 + 3\,\tau_4 - \rho_1 -
3\,\rho_2 \right) }{4\,{m_b}^3}
\,  \\
C_0^1 = &+ &\frac{ \Lambda}{{m_b}} - \frac{ \left(  \lambda_1 + 3\,  \lambda_2
- 2 \, n \,  {\Lambda}^2\right) }{2\,{m_b}^2}
+ \frac{ {n} \, \Lambda \,\left( \left(n  -1 \right) \, {\Lambda }^2 -
2\,\left( \lambda_1 + 3\, \lambda_2 \right) \right) }{2\,{m_b}^3} \nonumber \\
&-& \frac{  \left( \tau_1 + 3\,\tau_2 + \tau_3 + 3\,\tau_4 - \rho_1 - 3\,\rho_2
\right)   }{4\,{m_b}^3}
\, \\
C_0^2 = &+& \frac{ {{\Lambda }^2}}{2 \,{m_b}^2}
+ \frac{\Lambda \,\left( n\,{{\Lambda }^2} - \lambda_1 - 3\, \lambda_2
\right)}{2 \,{m_b}^3}
\, \\
C_0^3 = &+& \frac{ {{\Lambda }^3}}{3\,{m_b}^3}
\, \\
C_1^1 =  &-& \frac{\Lambda }{m_b} - \frac{ \left( 2\,\left( n -1  \right)
\,{{\Lambda }^2} + \lambda_1 + 3\, \lambda_2 \right) } {2\,{m_b}^2}
- \frac{ \left(n -1  \right) \,\Lambda \, \left( \left(n -2 \right) \,{{\Lambda
}^2} - 2\,\left( \lambda_1 + 3\, \lambda_2 \right)  \right)}{2\,{m_b}^3}
\nonumber \\
&+&  \frac{\left( \tau_1 + 3\,\tau_2 + \tau_3 + 3\,\tau_4 - \rho_1 -  3\,
\rho_2 \right) }{4\,{m_b}^3}
\, \\
C_1^2 = &-& \frac{ \, {{\Lambda }^2}}{{m_b}^2}
- \frac{2 \,\Lambda \,\left( \left(n  -1 \right) \,{{\Lambda }^2} - \lambda_1 -
3\,\lambda_2 \right) }{{m_b}^3}
\, \\
C_1^3 = &-& \frac{ {{\Lambda }^3}  }{ {m_b}^3}
\, \\
C_2^2 = & + & \frac{{{\Lambda }^2}}{{m_b}^2} +
\frac{\Lambda \,\left( \left(n -2  \right) \,{{\Lambda }^2} - \lambda_1 -
3\,\lambda_2 \right) }{{m_b}^3}
\, \\
C_2^3 = &+&\frac{{{\Lambda }^3}}{ {m_b}^3}
\, \\
C_3^3 = &-& \frac{  {{\Lambda }^3} }{3\,{m_b}^3}
\end{eqnarray}

\newpage

\section{{\bf Hadronic invariant Mass Moments for Fit}}
\renewcommand{\baselinestretch}{0.75}
{\normalsize{
\begin{center}
\vspace{0.3cm}
\begin{tabular}{|c||cccccccccccccc|}
\hline \hline
\, $\elmin$ \, & \, $S_{1/2}^0$ \,& \,$\Lambda_{1S}$ \,& \, $\Lambda_{1S}^2$
\,& \, $\Lambda_{1S}^3$ \,& \, $\lambda_1$ \,& \, $\lambda_2$ \,&  $\lambda_1
\, \Lambda_{1S}$
&  $\lambda_2 \, \Lambda_{1S}$ & \, $\rho_1$ \,& \, $\rho_2$ \,& \, $\tau_1$
\,& \, $\tau_2$ \,& \, $\tau_3$ \,& \, $\tau_4$ \,  \\
\hline \hline
0   & 2.1799 & 0.411 & 0.133 & 0.07 & 0.39 & -0.05 & 0.07 & 0.06 & 0.35 & -0.11
& 0.10 & 0.11 & 0.07 & 0.03 \\
0.5 & 2.1771 & 0.406 & 0.131 & 0.07 & 0.39 & -0.04 & 0.07 & 0.07 & 0.35 & -0.11
& 0.10 & 0.11 & 0.07 & 0.03  \\
0.7 & 2.1735 & 0.399 & 0.129 & 0.07 & 0.40 & -0.03 & 0.07 & 0.07 & 0.35 & -0.11
& 0.10 & 0.11 & 0.07 & 0.04 \\
0.9 & 2.1685 & 0.388 & 0.125 & 0.07 & 0.40 & -0.01 & 0.08 & 0.08 & 0.37 & -0.10
& 0.10 & 0.12 & 0.08 & 0.04 \\
1.1 & 2.1625 & 0.375 & 0.119 & 0.06 & 0.42 & 0.02 & 0.09 & 0.10 & 0.39 & -0.09
& 0.10 & 0.13 & 0.08 & 0.05 \\
1.3 & 2.1565 & 0.362 & 0.112 & 0.06 & 0.44 & 0.07 & 0.10 & 0.12 & 0.43 & -0.08
& 0.11 & 0.14 & 0.08 & 0.06 \\
1.5 & 2.1522 & 0.350 & 0.104 & 0.06 & 0.50 & 0.12 & 0.12 & 0.14 & 0.51 & -0.05
& 0.12 & 0.17 & 0.08 & 0.07  \\
\hline
\end{tabular}
\end{center}
}}
\begin{center}
TABLE 9: Coefficients of nonperturbative parameters for $S_{1/2}= S[0.5,
\elmin,0,\elmin]$.
\end{center}
\vspace{0.25cm}
\begin{center}
\renewcommand{\baselinestretch}{0.75}
{\normalsize{
\begin{tabular}{|c||cc|cc|cc|}
\hline \hline &
\multicolumn{2}{|c|}{ \rm{1S ${\alpha_s}^2$ Contribution}} &
\multicolumn{2}{|c|}{ \rm{ ${\alpha_s}$ Contribution} } &
\multicolumn{2}{|c|}{ \rm{Combined $O \left(\epsilon \right)$}}
\\ \hline
\, $\elmin$ \, & \,  $\epsilon$ \, & \,$\Lambda_{1S}$  $\epsilon $ \, &
 \,  $ \epsilon$ \, & \,$\Lambda_{1S}$  $\epsilon $ \, &  \,   $\epsilon$ \, &
\, $\Lambda_{1S}$  $\epsilon $ \, \\
\hline \hline
0 & -0.003 & 0.023 & 0.026 & 0.006 & 0.023  & 0.029  \\
0.5 & -0.003 & 0.022 &  0.031 & 0.014 & 0.027  & 0.037  \\
0.7 & -0.003 & 0.022 &  0.031 & 0.017 & 0.028  & 0.039  \\
0.9 & -0.003 & 0.021 &  0.031 & 0.020 & 0.027  & 0.041  \\
1.1 & -0.003 & 0.020 &  0.030 & 0.021 & 0.027  & 0.041  \\
1.3 & -0.004 & 0.019 &  0.029 & 0.023 & 0.025  & 0.042  \\
1.5 & -0.004 & 0.018 &  0.028 & 0.025 & 0.024  & 0.043  \\
\hline
\end{tabular}
}}
\end{center}
\begin{center}
TABLE 10: Coefficients of perturbative parameters for $S_{1/2}= S[0.5,
\elmin,0,\elmin]$.
\end{center}
\vspace{0.25cm}

\begin{center}
\renewcommand{\baselinestretch}{0.75}
{\normalsize{
\begin{tabular}{|c||cccccccccccccc|}
\hline \hline
\, $\elmin$ \, & \, $S_{3/2}^0$ \,& \,$\Lambda_{1S}$ \,& \, $\Lambda_{1S}^2$
\,& \, $\Lambda_{1S}^3$ \,& \, $\lambda_1$ \,& \, $\lambda_2$ \,&  $\lambda_1
\, \Lambda_{1S}$
&  $\lambda_2 \, \Lambda_{1S}$ & \, $\rho_1$ \,& \, $\rho_2$ \,& \, $\tau_1$
\,& \, $\tau_2$ \,& \, $\tau_3$ \,& \, $\tau_4$ \,  \\
\hline \hline
0   & 10.286 & 5.377 & 1.833 & 0.56 & 3.72 & -1.00 & 1.88 & 0.21 & 0.63 & -0.72
& 0.76 & 0.21 & 0.85 & 0.50 \\
0.5 & 10.249 & 5.296 & 1.801 & 0.55 & 3.77 & -0.87 & 1.90 & 0.30 & 0.67 & -0.74
& 0.77 & 0.24 & 0.86 & 0.51 \\
0.7 & 10.200 & 5.189 & 1.757 & 0.54 & 3.85 & -0.67 & 1.95 & 0.42 & 0.72 & -0.77
& 0.78 & 0.29 & 0.87 & 0.55 \\
0.9 & 10.133 & 5.040 & 1.689 & 0.53 & 4.00 & -0.35 & 2.04 & 0.63 & 0.81 & -0.81
& 0.81 & 0.37 & 0.88 & 0.60\\
1.1 & 10.054 & 4.859 & 1.597 & 0.51 & 4.25 & 0.09 & 2.19 & 0.92 & 0.97 & -0.84
& 0.85 & 0.50 & 0.91 & 0.67 \\
1.3 & 9.976 & 4.669 & 1.485 & 0.48 & 4.66 & 0.68 & 2.46 & 1.32 & 1.22 & -0.84 &
0.92 & 0.72 & 0.94 & 0.76 \\
1.5 & 9.919 & 4.515 & 1.366 & 0.44 & 5.40 & 1.43 & 2.96 & 1.83 & 1.69 & -0.72 &
1.05 & 1.10 & 0.98 & 0.89  \\
\hline
\end{tabular}
}}
\end{center}
\begin{center}
TABLE 11: Coefficients of nonperturbative parameters for $S_{3/2} = S[1.5,
\elmin,0,\elmin]$.
\end{center}
\newpage
\vspace{0.25cm}
\begin{center}
\renewcommand{\baselinestretch}{0.75}
{\normalsize{
\begin{tabular}{|c||cc|cc|cc|}
\hline \hline &
\multicolumn{2}{|c|}{ \rm{1S ${\alpha_s}^2$ Contribution}} &
\multicolumn{2}{|c|}{ \rm{ ${\alpha_s}$ Contribution} } &
\multicolumn{2}{|c|}{ \rm{Combined $O \left(\epsilon \right)$}}
\\ \hline
\, $\elmin$ \, & \,  $\epsilon$ \, & \,$\Lambda_{1S}$  $\epsilon $ \, &
 \,  $ \epsilon$ \, & \,$\Lambda_{1S}$  $\epsilon $ \, &  \,   $\epsilon$ \, &
\, $\Lambda_{1S}$  $\epsilon $ \, \\
\hline \hline
0 & -0.041 & 0.264 & 0.603 & 0.402 & 0.562  & 0.667  \\
0.5 & -0.041 & 0.261 &  0.483 & 0.225 & 0.443  & 0.485  \\
0.7 & -0.041 & 0.255 &  0.458 & 0.215 & 0.417  & 0.470  \\
0.9 & -0.041 & 0.248 &  0.432 & 0.216 & 0.391  & 0.464  \\
1.1 & -0.042 & 0.239 &  0.405 & 0.222 & 0.363  & 0.462  \\
1.3 & -0.045 & 0.229 &  0.380 & 0.236 & 0.335  & 0.465  \\
1.5 & -0.053 & 0.215 &  0.363 & 0.263 & 0.310  & 0.478  \\
\hline
\end{tabular}
}}\\
\end{center}
\begin{center}
TABLE 12: Coefficients of perturbative parameters for $S_{3/2} = S[1.5,
\elmin,0,\elmin]$.
\end{center}
\vspace{0.25cm}

\begin{center}
\renewcommand{\baselinestretch}{0.75}
{\normalsize{
\begin{tabular}{|c||cccccccccccccc|}
\hline \hline
 $\elmin$  & \, $S_{2a}^0$ \,& \,$\Lambda_{1S}$ \,& \, $\Lambda_{1S}^2$ \,& \,
$\Lambda_{1S}^3$ \,& \, $\lambda_1$ \,& \, $\lambda_2$ \,&  $\lambda_1 \,
\Lambda_{1S}$
&  $\lambda_2 \, \Lambda_{1S}$ & \, $\rho_1$ \,& \, $\rho_2$ \,& \, $\tau_1$
\,& \, $\tau_2$ \,& \, $\tau_3$ \,& \, $\tau_4$ \,  \\
\hline \hline
0   & 22.297 & 15.253 & 5.984 & 1.66 & 7.93 & -1.88 & 5.00 & 1.62 & -0.89 &
-1.12 & 1.47 & -0.89 & 2.23 & 1.41 \\
0.5 & 22.189 & 15.00 & 5.856 & 1.64 & 8.11 & -1.50 & 5.10 & 1.89 & -0.86 &
-1.25 & 1.50 & -0.79 & 2.25 & 1.47 \\
0.7 & 22.048 & 14.675 & 5.685 & 1.60 & 8.39 & -0.94 & 5.28 & 2.28 & -0.80 &
-1.41 & 1.55 & -0.63 & 2.28 & 1.55 \\
0.9 & 21.857 & 14.222 & 5.438 & 1.54 & 8.88 & -0.06 & 5.60 & 2.91 & -0.69 &
-1.63 & 1.63 & -0.38 & 2.32 & 1.68 \\
1.1 & 21.634 & 13.684 & 5.120 & 1.46 & 9.65 &  1.18 & 6.13 & 3.82 & -0.50 &
-1.87 & 1.76 & 0.01 & 2.38 & 1.87 \\
1.3 & 21.413 & 13.122 & 4.749 & 1.36 & 10.90 & 2.82 & 7.04 & 5.03 & -0.16 &
-2.05 & 1.97 & 0.63 & 2.46 & 2.12 \\
1.5 & 21.252 & 12.671 & 4.378 & 1.24 & 13.07 & 4.93 & 8.69 & 6.61 & 0.48 &
-2.00 & 2.30 & 1.64 & 2.57 & 2.44 \\
\hline
\end{tabular}
}}\\
\end{center}
\begin{center}
TABLE 13: Coefficients of nonperturbative parameters for $S_{2a} = S[2,
\elmin,0,\elmin]$.
\end{center}
\vspace{0.25cm}
\begin{center}
\renewcommand{\baselinestretch}{0.75}
{\normalsize{
\begin{tabular}{|c||cc|cc|cc|}
\hline \hline &
\multicolumn{2}{|c|}{ \rm{1S ${\alpha_s}^2$ Contribution}} &
\multicolumn{2}{|c|}{ \rm{ ${\alpha_s}$ Contribution} } &
\multicolumn{2}{|c|}{ \rm{Combined $O \left(\epsilon \right)$}}
\\ \hline
\, $\elmin$ \, & \,  $\epsilon$ \, & \,$\Lambda_{1S}$  $\epsilon $ \, &
 \,  $ \epsilon$ \, & \,$\Lambda_{1S}$  $\epsilon $ \, &  \,   $\epsilon$ \, &
\, $\Lambda_{1S}$  $\epsilon $ \, \\
\hline \hline
0 & -0.104 & 0.689 & 1.929 & 1.367 & 1.826  & 2.045  \\
0.5 & -0.103 & 0.679 &  1.425 & 0.596 & 1.321  & 1.275  \\
0.7 & -0.103 & 0.666 &  1.322 & 0.536 & 1.220  & 1.202  \\
0.9 & -0.104 & 0.647 &  1.227 & 0.526 & 1.124  & 1.173  \\
1.1 & -0.107 & 0.624 &  1.135 & 0.541 & 1.028  & 1.164  \\
1.3 & -0.115 & 0.596 &  1.051 & 0.578 & 0.937  & 1.174  \\
1.5 & -0.135 & 0.561 &  0.995 & 0.656 & 0.860  & 1.217  \\
\hline
\end{tabular}
}}
\\
\end{center}
\begin{center}
TABLE 14: Coefficients of perturbative parameters for $S_{2a} \left(E_0
\right)$.
\end{center}
\section{{\bf Lepton Energy Moments for Fit}}

Lepton energy moments appropriate for extracting $ \Lambda_{1S} $ and $
\lambda1 $ from previous work
\cite{christianmike} in terms of the inverse upsilon mass expansion and
$\bar{\Lambda}_{1S} \equiv \frac{m_{\Upsilon}}{2} - m_b^{\rm 1S}\, $definition
of
Eq.\ (\ref{defnlam1s}) via $\Lambda_{1S}^{\rm lepton} = \left(\bar{m}_B  -
\frac{m_{\Upsilon}}{2} \right) + \Lambda_{1S}$ are as follows.
The general moment is defined for the lepton spectrum in an identical fashion
to the
general moment for the hadronic invariant mass spectrum.

\begin{eqnarray}\label{momentdefinitlep}
R [n,E_{\ell_1},m,E_{\ell_2}] = \frac{\int_{E_{\ell_1}}^{E_\ell^{\rm max}}
E_\ell^n \frac{d\Gamma}{d E_\ell}d E_\ell}{\int_{E_{\ell_2}}^{E_\ell^{\rm max}}
\hat{E}_{\ell}^m
\frac{d\Gamma}{d E_\ell}d E_\ell}\,,
\end{eqnarray}

\begin{center}
\vspace{0.25cm}
\renewcommand{\baselinestretch}{0.75}
{\normalsize{
\begin{tabular}{|c||cccccccccccccc|}
\hline \hline
 $\elmin$  & \, $R_1^0$ \,& \,$\Lambda_{1S}$ \,& \, $\Lambda_{1S}^2$ & \,
$\Lambda_{1S}^3$ & \, $\lambda_1$ & \, $\lambda_2$ \,&  $\lambda_1 \,
\Lambda_{1S}$
&  $\lambda_2 \, \Lambda_{1S}$ & \, $\rho_1$ \,& \, $\rho_2$ \,& \, $\tau_1$
\,& \, $\tau_2$ \,& \, $\tau_3$ \,& \, $\tau_4$ \,  \\
\hline \hline
0   & 1.3920 & -0.075 & -0.02 & 0.00 & -0.10 & -0.21 & -0.04 & -0.05 & -0.03 &
0.01 & -0.02 & -0.01 & -0.03 & -0.03 \\
0.5 & 1.4216 & -0.074 & -0.02 & 0.00 & -0.10 & -0.21 & -0.04 & -0.05 & -0.03 &
0.01 & -0.02 & -0.01 & -0.03 & -0.03 \\
0.7 & 1.4611 & -0.073 & -0.02 & 0.00 & -0.10 & -0.20 & -0.04 & -0.05 & -0.03 &
0.01 & -0.02 & -0.02 & -0.03 & -0.03 \\
0.9 & 1.5173 & -0.073 & -0.02 & 0.00 & -0.10 & -0.20 & -0.04 & -0.05 & -0.04 &
0.01 & -0.02 & -0.02 & -0.03 & -0.03 \\
1.1 & 1.5884 & -0.073 & -0.02 & 0.00 & -0.10 & -0.19 & -0.04 & -0.05 & -0.04 &
0.00 & -0.02 & -0.02 & -0.02 & -0.03 \\
1.3 & 1.6724 & -0.074 & -0.02 & 0.00 & -0.10 & -0.18 & -0.04 & -0.05 & -0.04 &
0.00 & -0.02 & -0.02 & -0.02 & -0.03 \\
1.5 & 1.7674 & -0.076 & -0.02 & 0.00 & -0.11 & -0.17 & -0.04 & -0.05 & -0.05 &
-0.01 & -0.02 & -0.03 & -0.02 & -0.03 \\
\hline
\end{tabular}
}}
\end{center}
\begin{center}
TABLE 15: Coefficients of the nonperturbative parameters for $R_{1} = R[1,
\elmin,0,\elmin]$. \\
\end{center}
\vspace{0.25cm}
\begin{center}
\renewcommand{\baselinestretch}{0.75}
{\normalsize{
\begin{tabular}{|c||ccc|ccc|ccc|}
\hline \hline &
\multicolumn{3}{|c|}{ \rm{ \,1S ${\alpha_s}^2$\, \,  \, \, ${\alpha_s}^3 \,
\beta_0 $  }} &
\multicolumn{3}{|c|}{ \rm{ \, \,  ${\alpha_s}$ \, \, \, \, ${\alpha_s}^2 \,
\beta_0 $ } } &
\multicolumn{3}{|c|}{ \rm{Combined Terms}}
\\ \hline
\, $\elmin$ \, & \, \,  $\epsilon$ \,\,  & $ \, \, \Lambda_{1S}$  $\epsilon $
\, \, &   $\epsilon^2_{BLM}$  &
   $ \epsilon$ \, & \,$\Lambda_{1S}$  $\epsilon $  &   $\epsilon^2_{BLM}$  & \,
$\epsilon$ \,  \,& \,  $\Lambda_{1S}$  $\epsilon $  &   $\epsilon^2_{BLM}$   \\
\hline \hline
0 & 0.004 & 0.001 & 0.006 & -0.001 & -0.001 & -0.002 & 0.003  & 0.000  & 0.003
\\
0.5 & 0.004 & 0.001 & 0.006 & -0.001 & -0.001 & -0.003 & 0.003  & 0.000  &
0.002 \\
0.7 & 0.004 & 0.001 & 0.006 & -0.001 & -0.001 & -0.004 & 0.002  & 0.000  &
0.002 \\
0.9 & 0.004 & 0.001 & 0.006 & -0.002 & -0.001 & - 0.005 & 0.002  & 0.000  &
0.001 \\
1.1 & 0.004 & 0.001 & 0.006 & -0.002 & -0.001 & -0.006 & 0.001  & 0.000  &
0.000 \\
1.3 & 0.004 & 0.001 & 0.006 & -0.003 & -0.001 & -0.007 & 0.001  & 0.000  &
-0.001 \\
1.5 & 0.004 & 0.001 & 0.006 & -0.003 & -0.001 & -0.007 & 0.001  & 0.000  &
-0.001 \\
\hline
\end{tabular}
}}
\end{center}
\begin{center}
TABLE 16: Coefficients of the perturbative parameters for $R_{1} \left(E_0
\right)$.
\end{center}
\vspace{0.25cm}

\begin{center}
\renewcommand{\baselinestretch}{0.75}
{\normalsize{
\begin{tabular}{|c||cccccccccccccc|}
\hline \hline
 $\elmin$  & \, $V_1^0$ \,& \,$\Lambda_{1S}$ \,&  $\Lambda_{1S}^2$ \,&
$\Lambda_{1S}^3$ \,&  $\lambda_1$ \,& \, $\lambda_2$ \,&  $\lambda_1 \,
\Lambda_{1S}$
&  $\lambda_2 \, \Lambda_{1S}$ & \, $\rho_1$ \,& \, $\rho_2$ \,& \, $\tau_1$
\,& \, $\tau_2$ \,& \, $\tau_3$ \,& \, $\tau_4$ \,  \\
\hline \hline
0   & 0.1804 & -0.032 & 0.00 & 0.00 & -0.05 & -0.07 & -0.01 & -0.01 & -0.04 & -
0.01 & -0.01 & -0.02 & -0.01 & -0.01 \\
0.5 & 0.1542 & -0.032 & 0.00 & 0.00 & -0.05 & -0.06 & -0.01 & -0.01 & -0.04 & -
0.01 & -0.01 & -0.02 & -0.01 & -0.01 \\
0.7 & 0.1280 & -0.030 & 0.00 & 0.00 & -0.05 & -0.06 & -0.01 & -0.01 & -0.04 & -
0.01 & -0.01 & -0.02 & -0.01 & -0.01 \\
0.9 & 0.0988 & -0.028 & 0.00 & 0.00 & -0.05 & -0.06 & -0.01 & -0.01 & -0.04 & -
0.01 & -0.01 & -0.02 & -0.01 & -0.01 \\
1.1 & 0.0705 & -0.025 & 0.00 & 0.00 & -0.05 & -0.05 & -0.01 & -0.01 & -0.04 & -
0.01 & -0.01 & -0.02 & -0.01 & -0.01 \\
1.3 & 0.0458 & -0.021 & 0.00 & 0.00 & -0.05 & -0.04 & -0.01 & -0.01 & -0.04 & -
0.01 & -0.01 & -0.02 & -0.01 & -0.01 \\
1.5 & 0.0261 & -0.017 & 0.00 & 0.00 & -0.04 & -0.03 & -0.01 & -0.02 & -0.04 & -
0.01 & -0.01 & -0.02 & -0.01 & -0.01 \\
\hline
\end{tabular}
}}
\end{center}
\begin{center}
\vspace{0.25cm}
Table 17: Coefficients of the nonperturbative parameters for variance $V_1 =
(R[2, \elmin,0,\elmin] - R[1, \elmin,0,\elmin]^2)$.
\end{center}
\newpage

\begin{center}
\vspace{0.25cm}
\renewcommand{\baselinestretch}{0.75}
{\normalsize{
\begin{tabular}{|c||ccc|ccc|ccc|}
\hline \hline &
\multicolumn{3}{|c|}{ \rm{ \,1S ${\alpha_s}^2$\, \,  \, \, ${\alpha_s}^3 \,
\beta_0 $  }} &
\multicolumn{3}{|c|}{ \rm{ \, \,   ${\alpha_s}$ \, \, \, \,  \, \,
${\alpha_s}^2 \, \beta_0 $ } } &
\multicolumn{3}{|c|}{ \rm{Combined Terms}}
\\ \hline
\, $\elmin$ \, & \, \,  $\epsilon$ \,\,  & $ \, \, \Lambda_{1S}$  $\epsilon $
\, \, &   $\epsilon^2_{BLM}$  &
   $ \epsilon$ \, & \,$\Lambda_{1S}$  $\epsilon $  &   $\epsilon^2_{BLM}$  & \,
$\epsilon$ \, & \, $\Lambda_{1S}$  $\epsilon $  &   $\epsilon^2_{BLM}$   \\
\hline \hline
0 & 0.002 & 0.000 & 0.003 & -0.002 & 0.000 & -0.002 & 0.000  & 0.000  & 0.001
\\
0.5 & 0.002 & 0.000 & 0.003 & -0.002 & -0.001 & -0.001 & 0.000  & 0.001  &
0.001 \\
0.7 & 0.002 & 0.000 & 0.003 & -0.002 & 0.000 & -0.001 & 0.000  & 0.000  & 0.002
\\
0.9 & 0.001 & 0.000 & 0.002 & -0.001 & 0.000 & - 0.001 & 0.000  & 0.000  &
0.002 \\
1.1 & 0.001 & 0.000 & 0.002 & -0.001 &  0.000 & -0.001 & 0.000  & 0.000  &
0.002 \\
1.3 & 0.001 & 0.000 & 0.002 & -0.001 & 0.000 & -0.001 & 0.000  & 0.000  & 0.001
\\
1.5 & 0.001 & 0.000 & 0.002 & -0.001 & 0.000 & 0.000 & 0.000  & 0.000  & 0.001
\\
\hline
\end{tabular}
}} \\
\end{center}
\begin{center}
TABLE 18: Coefficients for lepton variance perturbative parameters. \\
\end{center}
\vspace{0.25cm}

\begin{center}
\renewcommand{\baselinestretch}{0.75}
{\normalsize{
\begin{tabular}{|c||cccccccccccccc|}
\hline \hline
\, $\elmin$ \, & \, $V_2^0$ \,& \,$\Lambda_{1S}$ \,& \, $\Lambda_{1S}^2$ \,& \,
$\Lambda_{1S}^3$ \,& \, $\lambda_1$ \,& \, $\lambda_2$ \,&  $\lambda_1 \,
\Lambda_{1S}$
&  $\lambda_2 \, \Lambda_{1S}$ & \, $\rho_1$ \,& \, $\rho_2$ \,& \, $\tau_1$
\,& \, $\tau_2$ \,& \, $\tau_3$ \,& \, $\tau_4$ \,  \\
\hline \hline
0   &  -0.0376 &  0.001 &  0.002 &  0.0  & -0.01 &  0.02 &  0.0   &  0.01  
& -0.03 & -0.01 &  0.0  & -0.01 &  0.0  &  0.0   \\
0.5 &  -0.0207 &  0.0   &  0.002 &  0.0  & -0.01 &  0.01 &  0.0   &  0.0   
& -0.03 & -0.01 &  0.0  & -0.01 &  0.0  &  0.0  \\
0.7 &  -0.0105 &  0.0   &  0.001 &  0.0  & -0.01 &  0.01 &  0.0   &  0.0 
& -0.03 & -0.01 &  0.0  & -0.01 &  0.0  &  0.0  \\
0.9 &  -0.0036 & -0.002 &  0.001 &  0.0  & -0.01 &  0.01 &  0.0   &  0.0 
& -0.03 & -0.01 &  0.0  & -0.01 &  0.0  &  0.0  \\
1.1 &  -0.0001 & -0.002 &  0.001 &  0.0  & -0.01 &  0.0  &  0.0   &  0.0 
& -0.02 & -0.01 &  0.0  & -0.01 &  0.0  &  0.0  \\
1.3 &   0.0009 & -0.002 &  0.0   &  0.0  & -0.01 &  0.0  &  0.0   &  0.0 
& -0.02 & -0.01 &  0.0  & -0.01 &  0.0  &  0.0  \\
1.5 &   0.0009 & -0.001 &  0.0   &  0.0  & -0.01 &  0.0  &  0.0   &  0.0  
& -0.02 &  0.0  &  0.0  &  0.0  &  0.0  &  0.0  
\\
\hline
\end{tabular}
}}\\
\end{center}
\begin{center}
TABLE 19: Coefficients for the nonperturbative parameters of $ V_2 = <(R[1,
\elmin,0,\elmin] - <R[1, \elmin,0,\elmin]>)^3>$.\\
\end{center}
\vspace{0.25cm}
\begin{center}
\renewcommand{\baselinestretch}{0.75}
{\normalsize{
\begin{tabular}{|c||ccc|ccc|ccc|}
\hline \hline &
\multicolumn{3}{|c|}{ \rm{ \,1S ${\alpha_s}^2$\, \,  \, \, ${\alpha_s}^3 \,
\beta_0 $  }} &
\multicolumn{3}{|c|}{ \rm{ \, \,  ${\alpha_s}$ \, \, \, \,  \, \, ${\alpha_s}^2
\, \beta_0 $ } } &
\multicolumn{3}{|c|}{ \rm{Combined Contributions}}
\\ \hline
\, $\elmin$ \, & \, \,  $\epsilon$ \,\,  & $ \, \, \Lambda_{1S}$  $\epsilon $
\, \, &   $\epsilon^2_{BLM}$  &
   $ \epsilon$ \, & \,$\Lambda_{1S}$  $\epsilon $  &   $\epsilon^2_{BLM}$  & \,
$\epsilon$ \, & \, $\Lambda_{1S}$  $\epsilon $  &   $\epsilon^2_{BLM}$   \\
\hline \hline
0 & 0.000 & 0.000 &  0.000 & 0.000 & 0.001 & 0.000 &  0.000  & 0.001  & 0.000
\\
0.5 & 0.000 & 0.000 &  0.000 & 0.001 & 0.000 & 0.000 &  0.001  & 0.000
 & 0.000 \\
0.7 & 0.000 & 0.000 &  0.000 & 0.000 & 0.000 & 0.000 &  0.000  & 0.000
 & 0.000 \\
0.9 & -0.008 & 0.000 &  0.000 & 0.007 & 0.001 & 0.000 & -0.001  & 0.001  &
 0.000 \\
1.1 & -0.007 & 0.000 &  0.000 & 0.005 &  0.000 & 0.000 & -0.002  & -0.001  &
 0.000 \\
1.3 & -0.006 & 0.000 &  0.000 & 0.005 & 0.000 & 0.000 & -0.001  & 0.001  &
 0.000 \\
1.5 & -0.005 & 0.000 &  0.000 & 0.003 & 0.000 & 0.000 & -0.002  & 0.000  &
 0.000 \\
\hline
\end{tabular}
}}
\end{center}
\begin{center}
TABLE 20: Coefficients for the perturbative parameters of  $ V_2$\\
\end{center}
\vspace{0.25cm}
\newpage

\begin{center}
 \renewcommand{\baselinestretch}{0.75}
{\normalsize{
\begin{tabular}{|c||cccccccccccccc|}
\hline \hline
& \,  \,& \,$\Lambda_{1S}$ \,& \, $\Lambda_{1S}^2$ \,& \, $\Lambda_{1S}^3$ \,&
\, $\lambda_1$ \,& \,
$\lambda_2$ \,&  $\lambda_1 \, \Lambda_{1S}$ &  $\lambda_2 \, \Lambda_{1S}$ &
\, $\rho_1$ \,& \, $\rho_2$ \,& \, $\tau_1$ \,& \,
$\tau_2$ \,& \, $\tau_3$ \,& \, $\tau_4$ \,  \\
\hline \hline
$D_1 $ & 0.5452 & 0.001 & -0.003 & -0.01 & 0.002 & -0.02 & 0.00 & -0.01 & 0.01
& 0.01 & 0.00 & 0.01 & 0.01 & 0.00 \\
$D_2 $ & 1.7626 & 0.014 & 0.014 & 0.00 & 0.001 & 0.09 & 0.01 & 0.05 & -0.01 &
-0.02 & 0.00 & -0.01 & 0.01 & 0.01 \\
$D_3 $ & 0.5215 & -0.011 & -0.009 & 0.00 & -0.002 & -0.04 & -0.01 & -0.03 &
0.01 & 0.01 & 0.00 & 0.00 & 0.00 & 0.00 \\
$D_4 $ & 0.6051 & -0.015 & -0.011 & -0.003 & -0.006 & -0.04 & -0.01 & -0.03 &
0.01 & 0.01 & 0.00 & 0.00 & 0.00 & 0.00 \\
\hline
\end{tabular}
}}
\end{center}
TABLE 21: Coefficients of the nonperturbative parameters of the  lepton energy
OPE testing Moments, where $D_1  = S\left[0.2,1.3,1,1 \right]$,
 $D_2 = S\left[0.8,1, 0.1, 1.3 \right]$ , $D_3  = S\left[0.7,1.6, 1.5 , 1.5
\right]$,
$D_d^4  = S\left[2.4,1, 1.9, 0.8 \right]$  and $D_e^5  = S\left[2.9,1.4, 2.2,
1.3 \right]$.

\vspace{0.25cm}
\begin{center}
\renewcommand{\baselinestretch}{0.75}
{\normalsize{
\begin{tabular}{|c||ccc|ccc|ccc|}
\hline \hline &
\multicolumn{3}{|c|}{ \rm{ \,1S ${\alpha_s}^2$\, \,  \, \, ${\alpha_s}^3 \,
\beta_0 $  }} &
\multicolumn{3}{|c|}{ \rm{ \, \,  ${\alpha_s}$ \, \, \, \,  \, \, ${\alpha_s}^2
\, \beta_0 $ } } &
\multicolumn{3}{|c|}{ \rm{Combined Contributions}}
\\ \hline
\, $\elmin$ \, & \, \,  $\epsilon$ \,\,  & $ \, \, \Lambda_{1S}$  $\epsilon $
\, \, &   $\epsilon^2_{BLM}$  &
   $ \epsilon$ \, & \,$\Lambda_{1S}$  $\epsilon $  &   $\epsilon^2_{BLM}$  & \,
$\epsilon$ \, & \, $\Lambda_{1S}$  $\epsilon $  &   $\epsilon^2_{BLM}$   \\
\hline \hline
$D_1 $ & 0.000 & 0.000 & 0.000 &  0.000 & 0.000 & 0.004 & 0.001 & 0.000 & 0.004
 \\
$D_2 $ & -0.001 & 0.000 & 0.000 &  -0.001 & -0.001 & -0.012 & -0.002 & -0.001 &
-0.012  \\
$D_3 $ & 0.000 & 0.000 & 0.000 &  0.000 & 0.000 & 0.003 & 0.000 & 0.000 & 0.003
 \\
$D_4 $ & 0.001 & 0.001 & 0.001 &  -0.001 & -0.001 & 0.003 & 0.000 & 0.000 &
0.004  \\
\hline
\end{tabular}
}} \\
\end{center}
TABLE 22: Coefficients of the perturbative parameters of the lepton energy OPE
Moments. \\

\vspace{0.25cm}
\begin{center}
\renewcommand{\baselinestretch}{0.75}
{\normalsize{
\begin{tabular}{|c|c|c|}
\hline \hline
\, Label &  \, Predicted Value \, & \, Measured Value \,\\
\hline \hline
$D_1 $ &  0.5459 $\pm 0.0001 \, \epsilon \, \pm 0.0010 ({\rm N.P.}) $ & - \\
$D_2 $ &   1.7585 $\pm 0.006 \, \epsilon \, \pm 0.0036 ({\rm N.P.}) $  & - \\
$D_3 $ &   0.5200 $\pm 0.0001 \, \epsilon \, \pm 0.0014 ({\rm N.P.}) $ & 0.5193
$\pm$ 0.0008 \\
$D_4 $ &  0.6053 $\pm 0.0002 \, \epsilon \, \pm 0.0018 ({\rm N.P.})$ & 0.6036
$\pm $ 0.0006 \\
\hline
\end{tabular}
}}
\end{center}
TABLE 23: Predictions and measurements for lepton energy OPE Testing Moments.\\


\begin{thebibliography}{99}
%
\bibitem{ope}
M.~A.~Shifman and M.~B.~Voloshin,
Sov. \ J.\ Nucl.\ Phys.\ {\bf 41}, 120 (1985);
J.~Chay, H.~Georgi and B.~Grinstein,
Phys.\ Lett. \ B{\bf 247}, 399 (1990).
%
\bibitem{opecorr1}
I.~I.~Bigi \textit{ et al.},Phys.\ Lett.\  B {\bf 293}, 430 (1992)
[(E) \textit{ibid.} B{\bf 297} 477 (1993)];
I.~I.~Bigi,\textit{ et al.},
Phys.\ Rev.\ Lett.\  {\bf 71}, 496 (1993).
%
\bibitem{opecorr2}
A.~V.~Manohar and M.~B.~Wise,
Phys.\ Rev.\ D {\bf 49}, 1310 (1994);
B.~Blok \textit{ et al.}
Phys.\ Rev.\ D {\bf 49}, 3356 (1994)[(E) \textit{ibid.} D{\bf 50} 3572 (1994)];
T.~Mannel,
Nucl.\ Phys. \ B {\bf 413} 396 (1994).
%
\bibitem{leptonmomentValoshin}
M.B.~Voloshin,
Phys.\ Rev.\ D {\bf 51}, 4934 (1995).
%
\bibitem{leptonmoment4}
M.~Gremm and I.~Stewart,
Phys.\ Rev.\ D {\bf 55}, 1226 (1997).
%
\bibitem{leptonmoment2}
M.~Gremm, A.~Kapustin, Z.~Ligeti and M.~B.~Wise,
Phys.\ Rev.\ Lett.\ {\bf 77}, 20 (1996).
%
\bibitem{hadronmoment2}
A.F.~Falk,  M.~Luke, M.J.~Savage,
Phys.\ Rev.\ D {\bf 53}, 2491 (1996).
%
\bibitem{hadronmoment3}
A.F.~Falk,  M.~Luke, M.J.~Savage,
Phys.\ Rev.\ D {\bf 53}, 6316 (1996).
%
\bibitem{hadron3rdorder}
M.~Gremm and A.~Kapustin,
Phys.\ Rev.\ D {\bf 55}, 6924 (1997).
%
%
\bibitem{Falkreview}
Adam F Falk,
Nucl. \ Phys. \ Proc. \ Suppl. \  {\bf{111}}, 3-13, (2002).
%
\bibitem{CLEO2001}
S.~Chen \textit{ et al.} (CLEO Collaboration),
Phys.\ Rev.\ Lett. {\bf 87}, 251807 (2001).
%
\bibitem{CLEO20012}
D.~Cronin-Hennessy \textit{ et al.} (CLEO Collaboration),
Phys.\ Rev.\ Lett {\bf 87}, 251808 (2001).
%
\bibitem{CLEO2002}
R.~Briere \textit{ et al.} (CLEO Collaboration),
hep-ex/0209024.
%
\bibitem{BABAR01}
B.~Aubert { et al.} (BABAR Collaboration),
hep-ex/0207084.
%
\bibitem{DELPHI02604}
DELPHI Collaboration, ICHEP 2002, 2002-071-CONF-605.
%
\bibitem{DELPHI02605}
DELPHI Collaboration, ICHEP 2002, 2002-070-CONF-604.
%
%
\bibitem{everyonebutme}
C. W. ~Bauer, Z. ~Ligeti, M. ~Luke, A.V. ~Manohar,
Phys.\ Rev.\ {\bf D67}, 054012 (2003).
%
\bibitem{russianfit}
M.~Battiglia et al.,
Phys.\ Lett.\  {\bf B556}, 41-49 (2003).
%
\bibitem{uraltsev}
I.~Gambino, N. ~Uraltsev,
hep-ph 0401063.
%
\bibitem{christianmike}
C.W.~Bauer,  M.~Trott,
Phys.\ Rev.\ D {\bf 67},014021 (2003).
%
\bibitem{lukebmass}
 Aida X. El-Khadra and M. Luke,
Ann. \ Rev.\ Nucl. \ Part.\ Sci.\ {\bf52}:201-251 (2002).
%
\bibitem{renorupsilon}
A.~Hoang, Z.~ Ligeti and A.V.~Manohar,
Phys.\ Rev.\ D {\bf 59}, 074017, (1999).
%
\bibitem{upmassmanohar}
A.~Hoang, Z.~ Ligeti and A.V.~Manohar,
Phys.\ Rev.\ Lett.\ {\bf 82}, 277, (1999);
A.~H.~Hoang and T.~Teubner,
Phys.\ Rev.\ D {\bf 60}, 114027 (1999)\,.
%
\bibitem{renormalon}
I.~I.~Bigi, M.~A.~Shifman, N.~G.~Uraltsev and A.~I.~Vainshtein,
Phys.\ Rev.\ D {\bf 50}, 2234 (1994).
$\,$ M.~Beneke and V.~M.~Braun,
Nucl.\ Phys.\ B {\bf 426}, 301 (1994).
%
%
\bibitem{luke_renormalon}
M.~E.~Luke, A.~V.~Manohar and M.~J.~Savage,
Phys.\ Rev.\ D {\bf 51}, 4924 (1995).
%
%
\bibitem{jezabekandkhun}
M.~Jezabek,  J.H..~Kuhn,
Nucl.\ Phys.\ B {\bf 320},20-44 (1989).
%
\bibitem{corbospec}
G.~Corbo,
Nucl.\ Phys.\ B {\bf 212},99-108 (1983).
%
\bibitem{hadronmoment1}
A.F.~Falk,  M.~Luke, M.J.~Savage,
Phys.\ Rev.\ D {\bf 53}, 2491 (1996).
%
\bibitem{Neubertbupaper}
F.~De ~Fazio and  M.~ Neubert,
{\bf JHEP06}, 017 (1999).
%
\bibitem{AliPietarinen}
A.~Ali and E. Pietarinen,
Nucl. \ Phys. \ B {\bf 154}, 519, (1979).
%
\bibitem{Nir}
Y.~Nir,
Phys.\ Lett.\ B {\bf 221}, 184 (1989).
%
%
\bibitem{hadronmoment4}
A.F.~Falk,  M.~Luke,
Phys.\ Rev.\ D {\bf 57}, 424 (1998).
%
%
\bibitem{twoloopmike}
M.~Luke, M.J.~Savage and M.B. Wise,
Phys.\ Lett.\ B {\bf 343}, 329 (1995).
%
\bibitem{twoloopsmith}
B.H ~Smith, M.B.~Voloshin,
Phys.\ Lett.\ B {\bf 340}, 176 (1994).
%
\bibitem{vacuumsat}
I.~I.~Bigi \textit{ et al.},
Phys.\ Rev.\ D {\bf 52}, 196, (1995).
%
\bibitem{valoshinbb}
Valoshin,~MB. 
Nucl. \ Phys. \ B\ {\bf 154}, 365 (1979).
%
\bibitem{leutwyler}
Leutwyler.~H, 
Phys. \ Lett. \ B\ {\bf 98}, 447 (1981).
%
\bibitem{Pineda}
Pineda.~A, 
Nucl. \ Phys. \ B\ {\bf 494}, 213 (1997).
%
\bibitem{RussianMoments}
Vainstein AI, Zakharov VI, Shifman MA.  
JETP. \ Lett. \ {\bf 27}, 55 (1978);
Shifman MA., et al.  
Phys. \ Lett. \ B \ {\bf 77}, 80 (1978);
Shifman MA., et al.  
Phys. \ Lett. \ B \ {\bf 147}, 385 (1979);
%
\bibitem{BrambillaMoments}
Brambilla.~N, Sumino Y, Vairo A. 
Phys. \ Lett. \ B\ {\bf 513}, 381 (2001).
%
\bibitem{benekesinger}
M.~Beneke and  A. Singer, 
Phys. \ Lett. \ B\ {\bf 471}, 233 (1999).
%
\bibitem{Yaouanc1}
A.Le ~Yaouanc et al.,
Phys.\ Lett.\  {\bf B488}, 153 (2000),
%
\bibitem{Yaouanc2}
A.Le ~Yaouanc et al.,
Phys.\ Rev.\  {\bf D62}, 074007 (2000),
%
\bibitem{Yaouanc3}
A.Le ~Yaouanc et al.,
Phys.\ Lett.\  {\bf B517}, 135 (2001),
%
\bibitem{Bigi}
I.I.~Bigi and N.~Uraltzev,
Int.\ J.\ Mod. \ Phys.  {\bf A16}, 5201 (2001).
%
\bibitem{Bigi2}
D.~Denson, I.I.~Bigi and T.~Mannel,
Nucl. \ Phys. \ B\ {\bf 665}, 367 (2003).
%
\bibitem{Isgur}
N.~Isgur,
Phys.\ Lett.\  B {\bf 448}, 111 (1999).
%
\bibitem{CLEO2002m}
A.~Mahmood \textit{ et al.} (CLEO Collaboration),
Phys. \ Rev. \ D\ {\bf 67}, 072001 (2003).
%
\bibitem{artuso}
M.~Artuso,
hep-ph/0312270.
\end{thebibliography}
\end{document}